\documentclass[sigconf]{acmart}
\AtBeginDocument{%
  }

\copyrightyear{2026}
\acmYear{2026}
\setcopyright{cc}
\setcctype{by}
\acmConference[C\&C '26]{Creativity and Cognition}{July 13--16, 2026}{London, United Kingdom}
\acmBooktitle{Creativity and Cognition (C\&C '26), July 13--16, 2026, London, United Kingdom}
\acmDOI{10.1145/3803784.3807524}
\acmISBN{979-8-4007-2583-8/2026/07}

\renewcommand{\textcolor}[2]{#2}
\usepackage{array}
\usepackage{enumitem}

\begin{document}


\title{Reviving Reflection-in-Action: Instilling Designerly Thinking in AI-Supported Ideation through Multimodal Prompting}


\author{Samangi Wadinambiarachchi}
\orcid{0000-0002-0953-5306}
\affiliation{%
  \institution{The University of Melbourne}
  \city{Melbourne}
  \country{Australia}}
\email{samangi.w@unimelb.edu.au}

\author{Jenny Waycott}
\orcid{0000-0002-4634-0532}
\affiliation{%
  \institution{The University of Melbourne}
  \city{Melbourne}
  \country{Australia}}
\email{jwaycott@unimelb.edu.au}

\author{Greg Wadley}
\orcid{0000-0003-4230-3440}
\affiliation{%
  \institution{The University of Melbourne}
  \city{Melbourne}
  \country{Australia}}
\email{greg.wadley@unimelb.edu.au}

\renewcommand{\shortauthors}{Wadinambiarachchi et al.}

\begin{abstract}
    Current AI-powered creativity support tools (AI-CSTs) primarily use text prompting to generate solution-oriented outputs. However, the potential value of multimodal prompting in designer-AI interaction, specifically the introduction of \textit{productive friction} to encourage iteration and reflection, has not been fully explored. To address this, we developed \texttt{SketchifAI}, a prototype AI-CST, and evaluated it with design students. In a mixed-methods, within-participants study, we examined how different input modalities (text, sketch, and sketch-plus-tags) affected design students' perceived ability to express their intent, their perception of creativity support, and their divergent thinking performance. \textcolor{teal}{Our preliminary findings suggest that the sketch modality tended to enhance fluency, with inconclusive evidence for differences in variety, originality, or quality compared to text modality}. Yet, paradoxically, \textcolor{teal}{participants} showed a strong preference for text prompting. We discuss how AI tools \textcolor{teal}{might} be designed to reintroduce reflection-through-sketching, ensuring that designer-AI interaction supports, rather than erodes, essential design skills in students.
\end{abstract}

\ccsdesc[500]{Human-centered computing~Empirical studies in HCI}

\keywords{Design, Sketching, Inspiration, Ideation, Generative AI}

\maketitle

\section{Introduction}

Imagine it is week three of a design studio --- a peak time for rapid ideation. A few years ago the room would have been filled with the scent of graphite and the rhythmic scratching of pencils. Students would have struggled with the blank page, some feeling paralysed before making their first pen-stroke. Yet that friction was \textcolor{teal}{productive}; it was the force that empowered students to translate abstract ideas into a physical form, enabling reflection. 

Today, the sketchbook has been mostly replaced by screens, and the rhythmic scratch of pencils is faint, replaced by the staccato clatter of keyboards. Many students no longer try to wrestle with visualising their ideas, instead racing to construct prompts that flood their screens with highly refined, photorealistic images.

Researchers argue that sketching functions as a thinking tool and an external memory~\cite{Goldschmidt2017ManualRelevant,buxton2010sketching}, enabling designers to experiment, to discover new insights, and to rapidly produce a stream of ideas that keeps pace with their cognition. Thumbnail sketches~\cite{buxton2010sketching} and Crazy 8s~\cite{Crazy82015design}\footnote{\url{https://designsprintkit.withgoogle.com/methodology/phase3-sketch/crazy-8s}} are some of the popular techniques used for brainstorming and ideation. A successful ideation activity encourages divergent thinking, enabling designers to explore diverse concepts without being constrained by the need to refine details now~\cite{Hammond2019SketchingCognition, gallagher2017sketching}. 

A plethora of Artificial Intelligence (AI) tools has recently emer\-ged. Vendors claim that AI can be used as a co-creative assistant, helping designers to kick-start the creative process when feeling stuck~\cite{Shen2025IdeationWeb:Co-Creation, Rezwana2023UserStudy}. However, while some studies suggest that AI can augment human creativity and promote divergent thinking~\cite{Hoggenmueller2023CreativeExplorations, Lewis2023AIxArtist:Block}, others indicate that AI tools could narrow creativity~\cite{Wadinambiarachchi2024TheThinking, Doshi2024}, and should be appropriated with caution. The diffusion of AI tools into design workflows has obviated the struggle to make the first penstroke in favour of the challenge of: \textit{``how do I explain my idea in the right words?''}. But professional designers report that text prompting is not necessarily the best way to describe their intent to AI, highlighting the risk that linguistic limitations might inhibit visual exploration~\cite{Park2024WeTools, Li2024UserIntelligence,Peng2024DesignPrompt:AI,Uusitalo2024ClayPractice}.

In the context of the design classroom, the research discussed above raises the question: Could the use of Generative AI (GenAI) in the early stages of design jeopardise the fostering of designerly thinking? Students risk shifting from being active creators to passive curators. This could impose the steep cost of \textit{cognitive atrophy}; the erosion of cognitive skills due to over-reliance and cognitive offloading on AI tools~\cite{qin2025timing} . While AI might enhance efficiency in that students can visualise refined ideas at speed, such `frictionless' design processes might erode the reflective thinking required to achieve true creative mastery. 

The risk of cognitive atrophy is further exacerbated by the UI of current GenAI tools, which force users to express their visual intent through a text-prompt, before exposing users to high-fidelity images. This immediate leap from idea to refined aesthetic can lead to design fixation~\cite{Wadinambiarachchi2024TheThinking}, defined as a ``blind adherence to a set of ideas or concepts limiting the output of conceptual design''~\cite[p.~1]{Jansson1991DesignFixation}. It can be difficult for users to deconstruct these polished visual outputs into the fluid, malleable concepts necessary for novel idea generation~\cite{Vasconcelos2016InspirationChallenges, Jansson1991DesignFixation}. 

Despite the rapid growth of investigation into Human-AI co-creativity, a gap remains in HCI literature regarding the tension between efficient automation and the cognitive effects of text-to-image workflows. For example,~\citet{Davis2025SketchAI:Design, Lin2025Inkspire:Sketching} explored how sketching can be used to mitigate users' difficulty in translating ideas into words;~\citet{Choi2025Expandora:Model} incorporated stepwise textual scaffolding to specify user intent through progressive prompting; while~\citet{Park2025ExploringTools} explored how annotation and scribbling can define creative intent. Collectively, these findings suggest that GenAI interfaces need to support diverse input modalities, allowing designers to adaptively switch between modalities across different stages of the process~\cite{Lin2025Inkspire:Sketching, Lee2024TheProcess}.

While these offer valuable insights, there remains a need to empirically examine how \textcolor{teal}{\textit{productive friction} --- deliberate resistance embedded in an interaction that creates moments of reflection and clarification~\cite{shin2026interrogating,Sheahan2024DesiginingwithFriction,cox2016design,Chen_Schmidt_2024}} --- influences reflective ideation, and the possible gap between participant perceptions and expert assessment of design ideas. We address this gap by investigating how design students communicate their creative intent to AI tools via different interaction modalities --- text input, pure sketching, and a hybrid of sketching with short descriptive tags. Through a within-participant study with nine design students ($N=9$), we \textcolor{teal}{explore} how effortful sketching and the deliberate generation of low-mid fidelity outputs impact inspiration retrieval in rapid ideation. We used a purposive sample, prioritising subjective depth over broad generalisability, and employed reflexive thematic analysis~\cite{Braun2022ThematicGuide} and Bayesian hierarchical modelling~\cite{RichardMcElreath2015StatisticalRethinking} to \textcolor{teal}{derive preliminary findings through rigorous mixed-methods} in understanding \textbf{\textit{how reintroducing sketching and imperfect stimuli as productive friction affects students' perceptions of creativity support, divergent thinking, and alignment with creative intent}}. Our study investigated the following research questions:

\begin{itemize} [leftmargin=*]
    \item \textbf{RQ1:} How do different input modalities (Text-only, Sketch-only, and SketchPlusTags) influence students' perceptions of their ability to express creative intent and the relevance of the AI-generated outcomes? 
    \item \textbf{RQ2:} How do these input modalities influence the level of creativity support perceived by students\footnote{As measured by the Creativity Support Index (CSI)~\cite{CherryQuantifyingIndex}.}? 
    \item \textbf{RQ3:} To what extent do these modalities influence students' divergent thinking performance during a rapid ideation task\footnote{Measured using expert ratings for fluency, variety, originality, and quality.}? 
\end{itemize}

\noindent We contribute to Designer-AI Interaction literature in three ways: 
\begin{itemize}[leftmargin=*]
    \item Empirical Contribution: We provide evidence that while the \texttt{Sketch} modality shows a trend towards enhanced fluency, the addition of tags (\texttt{SketchPlusTags}) \textcolor{teal}{may} constrain Originality, despite serving as a clarity anchor for intent. This reveals a paradox where students preferred text-based prompting despite the potential generative benefits of sketching.
    \item Implications for Design: We propose dialogic interaction that prioritises critical reflection over seamlessness. This can ensure that GenAI supports \textit{designerly thinking} by positioning the interface as a site for reflection rather than just an inspiration generator.
    \item \textcolor{teal}{We introduce a research probe: a multimodal prompting interface (Text, Sketch, SketchPlusTags), to investigate the role of productive friction in AI-supported ideation. A key design decision implemented in this artefact is that the \texttt{Sketch} condition allows no text-input from the user, requiring users to externalise their intent purely through sketching} (see Section~\ref{sec:SketchifAI}).
\end{itemize}

\section{Background}

\subsection{Sketching as a Promoter of Designerly Thinking}

Sketching is widely practised in design and is considered essential to designerly activity, as it enables designers to rapidly experiment and generate insights~\cite{cross2023design}. Research indicates that sketching performs two primary roles in idea generation: supporting introspective inspiration searches~\cite{Goldschmidt2017ManualRelevant, cross2023design} and enabling feedback during the generative process. This cognitive process is best understood through Sch{\"o}n’s framework of reflection-in-action~\cite{schon2017reflective,chiapello2022s}. In this view, design is not seen as a linear execution of a pre-formed idea but a ``conversation with the materials.'' As the designer sketches, the ambiguity of the stroke talks back, triggering new interpretations and unintended discoveries. This iterative exercise is foundational to divergent thinking~\cite{gallagher2017sketching}, where fluency (the volume of ideas) serves as a precursor to variety and originality~\cite{Weiss2022IsOriginality, Dumas2014UnderstandingPerspective}. By exploring a larger conceptual space through rapid, low-fidelity sketching, designers avoid premature closure and reach more innovative solutions~\cite{Boden2009ConceptualSpaces, Jansson1991DesignFixation}.

The broad uptake of digital design tools has led researchers to highlight a risk of decline in sketching ability and reflection -- particularly among emerging designers, who may be more attuned to using graphics software than pencil and paper. More recently, the emergence of AI-powered design tools creates two additional risks. First, designers may find it challenging to communicate visual ideas through text prompts; designers are forced to translate visual intent into words -- a process that bypasses the spatial reasoning inherent in sketching~\cite{Uusitalo2024ClayPractice, Takaffoli2024GenerativeIndustry}, which remain the primary mode of interaction with image generation tools. Second, AI tools immediately produce high-fidelity visual outputs, narrowing the opportunity for designers to engage in a traditional design-thinking process with low-fidelity sketches ~\cite{Wadinambiarachchi2024TheThinking}.

Because current GenAI tools prioritise seamlessness, they introduce a risk of cognitive atrophy.~\citet{cox2016design} argue that productive friction can ``disrupt `mindless' automatic interactions, prompting moments of reflection and more `mindful' interaction''~[p.~1]. Furthermore, as \citet{Wadinambiarachchi2024TheThinking} notes, the immediate exposure of high-fidelity GenAI outputs can lead to design fixation, when used in rapid ideation tasks, narrowing idea exploration prematurely and fixating designers on AI’s initial interpretation. This offloading of the ideation process to a black-box limits the designer's opportunity to exercise reflective thinking, potentially leading to a long-term erosion of the skills required for independent ideation. Our work investigates how reintroducing sketching and imperfect GenAI images as deliberate forms of interaction friction can restore the reflective exercise essential to designerly thinking.

\subsection{AI as a Tool for Generating Inspiration}

Currently, there is interest in whether AI can provide creativity support and facilitate meaningful collaboration in design. Early explorations, such as the Creative Sketching Partner (CSP)~\cite{Davis2019CreativeCreativity}, demonstrated that AI could reduce design fixation by providing unexpected visual stimuli that disrupts a designer’s narrow focus. Since then, the field has moved toward \textit{mixed-initiative} systems where the AI acts as a co-creator rather than a passive tool~\cite{Deterding2017Mixed-InitiativeInterfaces}.

A growing literature explores interaction beyond prompting, such as through sketch-based AI interfaces. Such work includes, SketchAI~\cite{Davis2025SketchAI:Design}, Inkspire~\cite{Lin2025Inkspire:Sketching} and ImaginationVellum~\cite{marquardt2025imaginationvellum} represent a shift toward allowing designers to maintain spatial control, with the latter providing an infinite canvas for progressive refinement -- a process that resonates with the iterative discovery described by~\citet{schon2017reflective}. Similarly, Objective Portrait~\cite{van2023objective} emphasises meta-cognition by reflecting the designer’s process back to them to stimulate new divergent paths. Studies of 3D sketching~\cite{Lee2024TheProcess} have shown that linguistic prompts are often insufficient for conveying complex volumes, necessitating a return to sketch modalities.

Most such research has used the approach of designing a prototype, and assessing its usability and user perceptions of creativity support~\cite{Choi2025Expandora:Model, Choi2024CreativeConnect:AI, Lin2025Inkspire:Sketching, Lee2024TheProcess}. For example, Choi et al.~\cite{Choi2024CreativeConnect:AI} designed a GenAI pipeline that recombines sketches and texts, finding that participants perceived this approach as more supportive of creativity than a text-only baseline. In a study with 12 participants, Lin et al.~\cite{Lin2025Inkspire:Sketching} compared their prototype, Inkspire, with ControlNet (a neural network architecture that guides pre-trained text-to-image diffusion models), noting that Inkspire promoted inspiration and user satisfaction. Similarly, Lee et al.~\cite{Lee2024TheProcess} highlighted the critical role of sketching in exploring ideas for 3D generative images.

However, these approaches primarily investigate user preferences and perceived support, often omitting objective measures of creative output. This leaves open the question of whether the tool materially impacts the design process, or whether the satisfaction reported by users merely reflects reduced effort, which we identify as a potential driver of cognitive atrophy. As discussed~\cite{cox2016design,jonsson2022cracking}, a tool that is too easy may bypass the productive friction required for deep designerly thinking. Without evaluating creative thinking performance through expert assessment, it remains unclear if these tools genuinely support the designer or simply automate the path to a high-fidelity, yet potentially fixated, result.

Our work contributes to research on AI-powered creativity support in two ways: first, by exploring how designers communicate intent through varied modalities (Text, Sketch, and SketchPlusTags); and second, by providing empirical evidence that bridges the gap between participant perceptions and expert assessment. By employing Bayesian hierarchical modelling, we move beyond simple satisfaction metrics to \textcolor{teal}{explore} how interaction friction influences divergent thinking and creative outcomes of design students.

\section{SketchifAI}
\label{sec:SketchifAI}
\begin{figure*}[tbp]
    \centering
    \includegraphics[width=.95\linewidth]{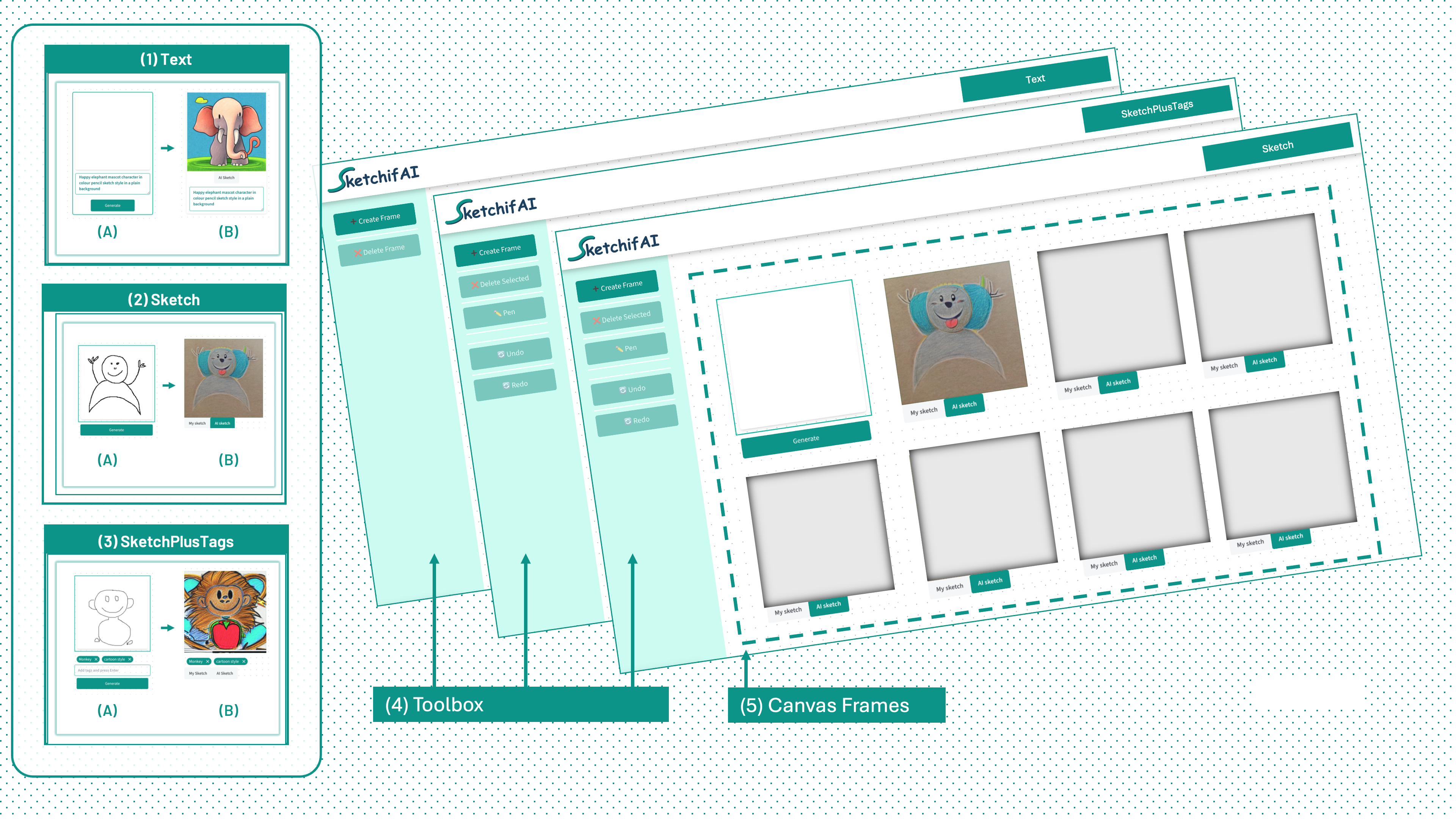}
    \caption{Interfaces of \textit{SketchifAI}. \textcolor{teal}{The left panel shows three input modalities: (1) Text, (2) Sketch, and (3) SketchPlusTags, where (A) shows the user's input and (B) shows the corresponding AI-generated sketch output. The right panel shows the application UI for each modality, stacked with Text (back), SketchPlusTags (middle), and Sketch (front), each comprising (4) a Toolbox for drawing controls and (5) Canvas Frames.}}
    \Description{This figure shows the interfaces of SketchifAI across two panels. The left panel presents three input modalities; each illustrated with an example. In the Text modality (1), a user enters a text prompt (A) and receives an AI-generated coloured sketch of an elephant mascot character (B). In the Sketch modality (2), a user provides a simple hand-drawn line sketch (A) and receives an AI-generated sketch of a bear-like mascot character (B). In the SketchPlusTags modality (3), a user provides a hand-drawn sketch accompanied by descriptive tags (A) and receives an AI-generated sketch of a monkey mascot character holding an apple (B). The right panel shows the full application UI in a stacked arrangement representing all three modalities, with Text at the back, SketchPlusTags in the middle, and Sketch at the front. Each UI instance includes a Toolbox (4) on the left with drawing controls and a Canvas Frames area (5) on the right displaying a grid of frames where participants' input sketches and AI-generated sketch outputs are shown side by side, labelled My Sketch and AI Sketch respectively. The Toolbox controls vary by modality: the Text modality includes Create Frame and Delete Frame, while the Sketch and SketchPlusTags modalities include Create Frame, Delete Selected, Pen, Undo, and Redo.}
    \label{fig:SketchifAI}
\end{figure*}

We created \texttt{SketchifAI} (see Figure~\ref{fig:SketchifAI}), a prototype that supports designers by generating inspiration stimuli in response to prompts entered through three distinct interfaces: \texttt{Text} (baseline), \texttt{Sketch}, and \texttt{SketchPlusTags}. We made a simple and consistent UI, with individual canvas frames tailored to support respective input modalities. 

\textcolor{teal}{While sketch-based interfaces have been explored in prior work, such as SketchAI~\cite{Davis2025SketchAI:Design}, Inkspire~\cite{Lin2025Inkspire:Sketching}, and CreativeConnect~\cite{Choi2024CreativeConnect:AI}, a key design decision is that neither \texttt{Sketch} nor \texttt{SketchPlusTags} condition requires a full text prompt. The back-end prompt automatically interprets the sketch outline, allowing the users to externalise their intent purely through sketching. In the \texttt{SketchPlusTags} condition, users can add short descriptive tags to guide the AI output, without needing to compose a full prompt}.

\subsection{System Walkthrough}

\subsubsection{Text Input Interface}

The \texttt{Text} interface (see Figure~\ref{fig:SketchifAI}-1) represents existing text-to-image tools. It allows users to create and manage canvas frames via a side panel. Each frame consists of a prompt input field, a generate button, and a display area for the resulting image. For example, a user might input: ``Happy elephant mascot character in coloured pencil sketch style in a plain background.'' Users could create multiple frames and keep generating images to generate inspiration stimuli for their work.

\subsubsection{Sketch Input Interface}

The \texttt{Sketch} interface (see Figure~\ref{fig:SketchifAI}-2) gives users a pen tool and drawing canvas with standard undo/redo functionality. After drawing a sketch, users can click ``generate'' to produce AI-interpreted variations of their sketch. A toggle feature allows users to switch between their original sketch and the AI-generated output (`my sketch' vs. `AI sketch'). This allows for iterative refinement and the exploration of new ideas based on the user's initial sketch outline.

\subsubsection{SketchPlusTags Input Interface}

The \texttt{SketchPlusTags} see Figure~\ref{fig:SketchifAI}-3) interface functions similarly to the sketch interface but with an additional text box for descriptive tags. Users generate images by combining visual sketches with short text descriptors-Tags. This hybrid approach enables the AI to produce more contextually accurate stimuli.

\subsection{Technical Implementation}

SketchifAI was developed as a web application (Next.js/Python) designed for use with a drawing tablet and stylus; for this study, we used the Huion Inspiroy H430P. For image generation, we utilised Stable Diffusion v1.5\footnote{\url{https://huggingface.co/stable-diffusion-v1-5/stable-diffusion-v1-5}} . To interpret user sketches we utilised ControlNet-Scribble\footnote{\url{https://huggingface.co/lllyasviel/sd-controlnet-scribble}}, which uses a neural network architecture to guide pre-trained text-to-image diffusion models using scribble images as additional input~\cite{Zhang2023AddingModels}. We selected this model over newer variants (such as SDXL) because it is lightweight, well-documented, and optimised for human-style scribbles. All sketches were preprocessed as $512 \times 512$ binary maps.

To integrate productive friction and maintain consistency, we utilised a back-end prompt for the \texttt{Sketch} and \texttt{SketchPlusTags} conditions:

\begin{itemize}[leftmargin=*, label={}]
    \item \texttt{Sketchy outline colour pencil drawing of a mascot character with new additional details to the sketch in plain background, HD+, HQ.}
\end{itemize}

This prompt ensured that all AI outputs had low-to-mid fidelity. This was in line with prior research which has shown that partially-complete or imperfect visual representations can stimulate more effective idea generation than completed ones~\cite{Cheng2014ADesigners,Youmans2014DesignColors, Alipour2016AStage}. By enforcing a \textit{sketchy} aesthetic, we aimed to force participants to actively interpret and build upon the AI's suggestions rather than passively accepting a polished, final solution. The system was hosted on a Google Cloud Platform\footnote{https://cloud.google.com} Linux-based virtual machine with an NVIDIA T4 GPU, achieving a generation latency of 7-10 seconds per image.

\section{Method}

We conducted a within-participant study to understand how different input modalities to AI affect: 1) ability to express intent to AI, 2) creativity, and 3) divergent thinking performance. Participants completed one mascot-design task under each input modality. To mitigate reduce carry-over and learning effects, condition exposure was randomised and design briefs were counterbalanced across the sample (see supplementary material-A for our Latin square design). This approach isolated the impact of interface design while controlling for task-specific influences. Our independent variable--\textit{input modality}--had three levels:

\begin{itemize}
    \item \texttt{(Baseline condition) Text}: Participants generated inspiration stimuli via text input; this mimics the current input interaction used in AI image generation. 
    \item \texttt{Sketch}: Participants generated inspiration stimuli via sketches.
    \item \texttt{SketchPlusTags}: Participants generated inspiration stimuli via sketches and small descriptive tags.
\end{itemize}


\subsection{Design Materials}

We asked participants to create mascot characters to serve as the logo and a key visual element for an app. (see Appendix~\ref{appendix_B} - for full briefs given to the participants.) We drew inspiration for our briefs from Ward's creature invention task~\cite{Kozbelt2007UnderstandingPredictors, Ward1994StructuredGeneration}, which directed participants to imagine and create aliens that lived on a different planet, and Choi et al.'s~\cite{Choi2025Expandora:Model} work on using mascots for design ideation tasks.

\begin{figure*}[tbp]
    \centering
    \includegraphics[width=.95\linewidth]{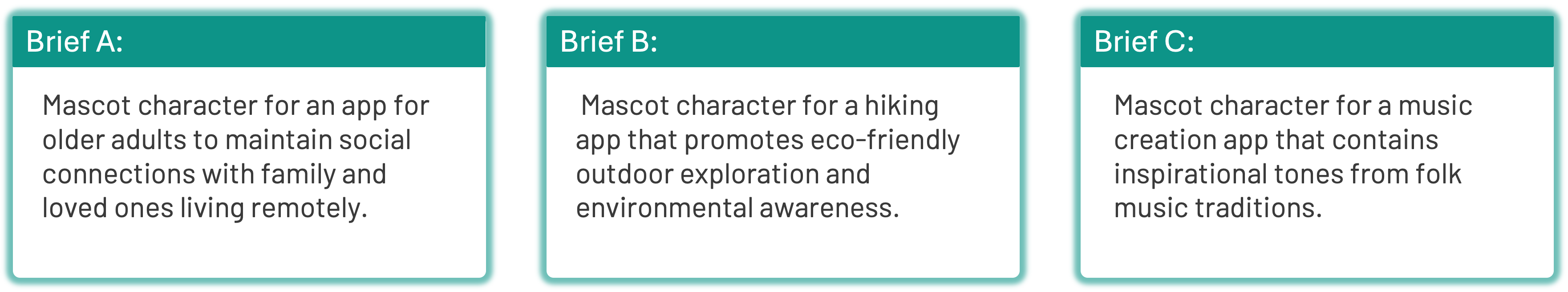}
    \caption{Overview of briefs provided to participants. \textcolor{teal}{Brief A (left), Brief B (middle), and Brief C (right).}}
    \Description{This figure shows three design brief topics arranged in a row. Each brief asks participants to design a mascot character for a different application: Brief A for an app helping older adults maintain social connections with family and loved ones living remotely; Brief B for a hiking app promoting eco-friendly outdoor exploration and environmental awareness; and Brief C for a music creation app incorporating inspirational tones from folk music traditions. The three briefs vary in target audience and domain — social connectivity, outdoor exploration, and folk music — while sharing the same task format: designing a mascot character.}
    \label{fig:Briefs}
\end{figure*}

The three briefs were designed to be abstract yet equivalent in complexity. The study received human research ethics approval from the University of Melbourne.

\subsection{Participants}

We recruited nine design students undertaking a Bachelor of Design or Master's degree in UX/UI, graphic, or a related design field, via university digital noticeboards, student clubs, social media, and word-of-mouth. Participants registered their interest in the study via an online form. Their responses were screened to ensure they were aged 18+ and had prior experience using AI tools, designing characters or avatars, and were able to sketch with a stylus on a drawing tablet. Eligible participants were invited via email. Participants were aged between 19 and 28 years (mean = 21, SD = 2.8). We ensured participants did not have a direct connection with the researchers involved in this study (e.g., student-teacher or supervisor-employee) to avoid performance anxiety and to maintain ethical integrity (see Appendix~\ref{appendix A} for participant information).

\subsection{Measures}

Participants first completed a brief demographic questionnaire (age, gender, design experience, and prior AI-tool use). To evaluate support for creativity, we administered a post-task questionnaire that included custom items alongside the Creativity Support Index (CSI)~\cite{CherryQuantifyingIndex} . Intent expression was measured specifically through two 7-point Likert scale questions (1 = Strongly disagree to 7 = Strongly agree) adapted from \citet{Choi2025Expandora:Model}:
\begin{enumerate}
    \item I was able to accurately express the direction of exploration I wanted to the system.
    \item I was able to clearly define the scope of exploration I wanted to the system.
\end{enumerate}

To evaluate our prototype, we used UMUX-Lite~\cite{UMUX-Lite}, a short instrument that correlates with System Usability Scale~\cite{BrookeSUSScale}.

\subsection{Procedure}
\begin{figure*}[tbp]
    \centering
    \includegraphics[width=.95\linewidth]{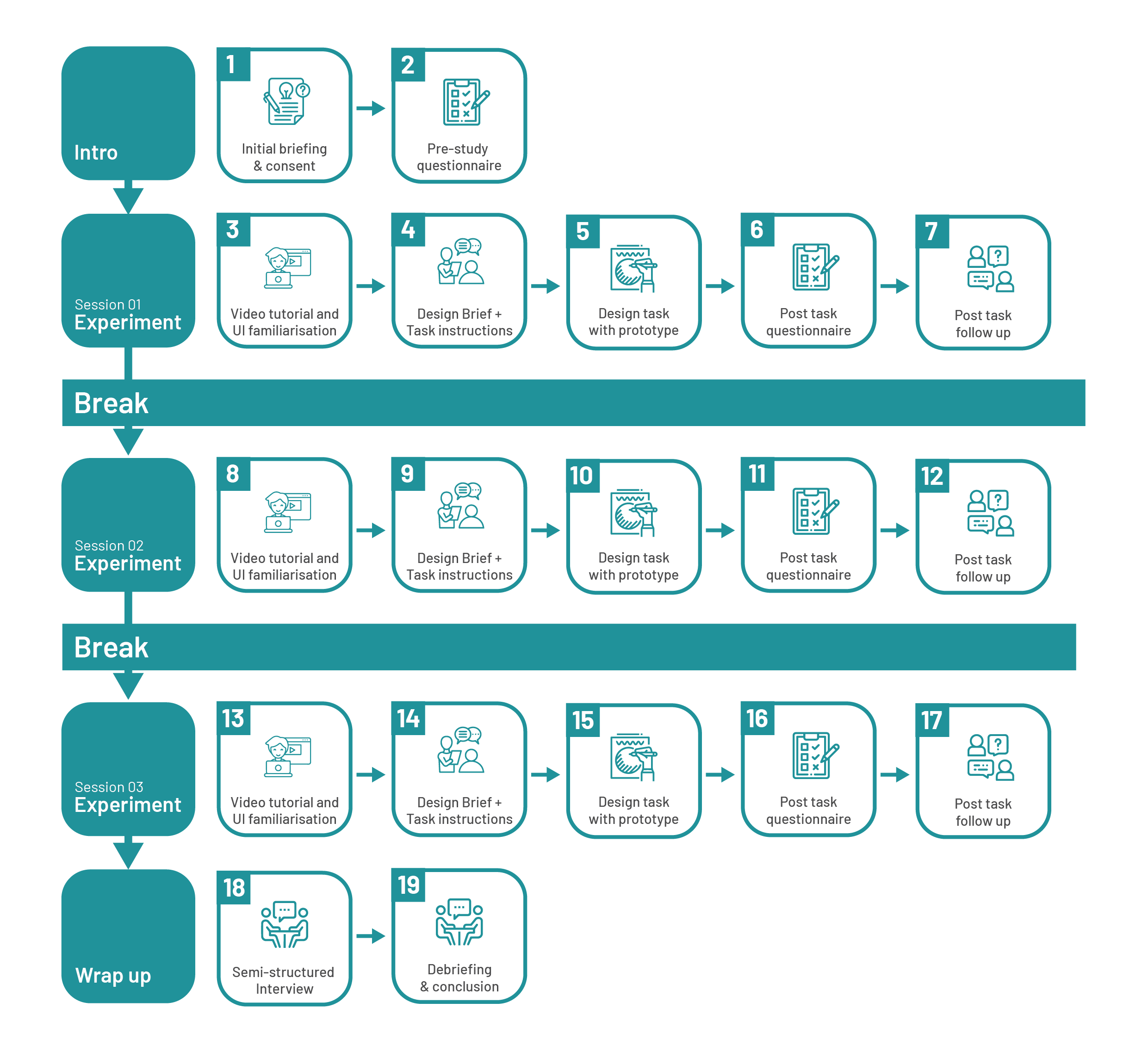}
    \caption{The overall experiment flow \textit{(1--2): Initial briefing, consent, questionnaire, (3--17): Main experimental sessions, (18--19) Semi-structured interview and debriefing.}}
    \Description{This figure shows a diagram of the overall experiment flow, organised into four phases. In the Intro phase, participants completed an initial briefing and consent form, followed by a pre-study questionnaire. Three experimental sessions followed, each separated by a break. Every session followed the same structure: a video tutorial and UI familiarisation, a design brief and task instructions, a design task using the prototype, a post-task questionnaire, and a short follow-up. In the Wrap Up phase, participants completed a semi-structured interview followed by a debriefing and conclusion.}
    \label{fig:procedure}
\end{figure*}

The study sessions were conducted in person in a UX research laboratory. Upon arrival, participants were given time to thoroughly read the project information sheet, and gave informed consent to participate (see Figure~\ref{fig:procedure}). Each experiment lasted~\textasciitilde 2hrs and 15mins. The first author conducted all sessions. 

In the \textbf{pre-study questionnaire}\footnote{(deployed via Qualtrics~\url{https://www.qualtrics.com})}, participants were automatically assigned a random, unique ID. The questionnaire collected participants' background information, including age, design experience, familiarity with AI tools, and confidence in sketching and prompting.

In the \textbf{main experimental sessions}, participants were first asked to watch a two-minute video tutorial about the prototype. They were then given three minutes to familiarise themselves with the prototype, during which they were encouraged to generate any images of their choice. Participants were given verbal instructions about the design briefs and allowed to ask questions to clarify any uncertainties. 

\begin{figure}[bp]
    \centering
    \includegraphics[width=.85\linewidth]{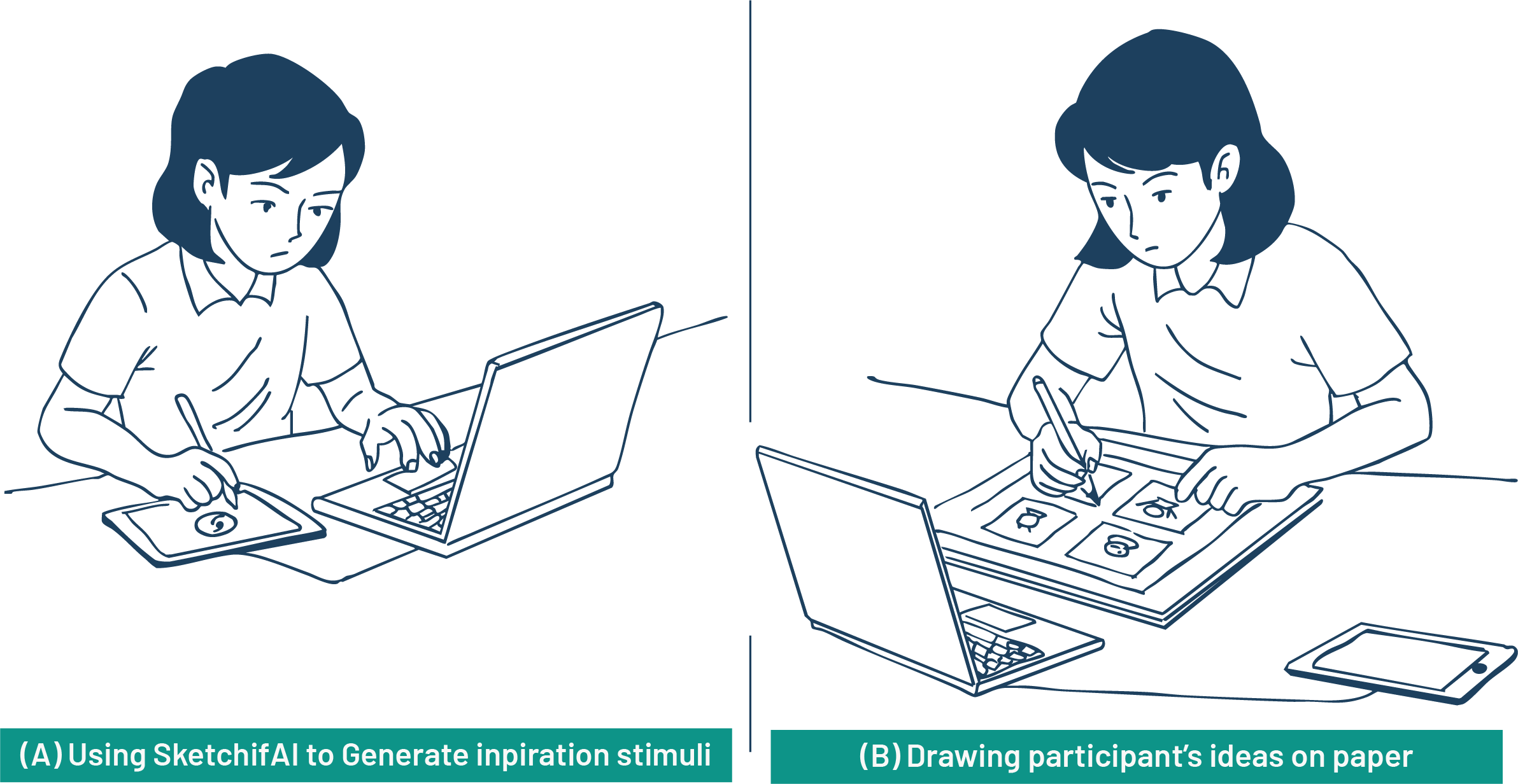}
    \caption{Participant activity workflow. During the study participants switched between (A) an inspiration phase where participants used \texttt{SketchifAI} to generate AI stimuli, and (B) sketch out their idea products on paper.}
    \Description{This figure shows two side-by-side illustrations depicting the two phases of participant activity during the study. In the left illustration (A), a participant is seated at a desk using a laptop and a drawing tablet, representing the inspiration phase in which participants used SketchifAI to generate inspiration stimuli. In the right illustration (B), the same participant is seated at a desk drawing in a paper sketchbook, with a laptop open in front of them and the drawing tablet set aside, representing the sketching phase in which participants sketched their design ideas by hand on paper. Note that in the Text condition, participants interacted with SketchifAI using only the keyboard, without the drawing tablet.}
    \label{fig:experiment}
\end{figure}

Participants were asked to complete three structured rapid ideation tasks. Each experimental session was 20 minutes long, to minimise participant fatigue. Prior studies have found this to be ideal for maintaining focus~\cite{Vasconcelos2016InspirationChallenges, Youmans2011TheFixation}. In each task, participants interacted with the prototype and then sketched their own ideas on paper. Hereafter we refer to these paper sketches as the participant's ``products'' (see Figure~\ref{fig:experiment}). 

Each participant then took part in a \textbf{post-session} questionnaire and short interview (see Figure~\ref{fig:procedure}). Participants took an eight-minute break between sessions. After the three design ideation sessions, we conducted a \textbf{semi-structured interview}, focusing on usability, creative engagement with the AI tools, and participants' expectations for interacting with such devices in the future. The participants were then debriefed, given the opportunity to ask questions, and given a summary of the study and prototype implementation. We thanked each participant with a \$50 AUD gift voucher. 

\subsection{Data Preparation}

We scanned all products created by participants and assigned a unique ID to each, before importing them to a Miro board\footnote{\url{https://miro.com}} for expert clustering and evaluations. \textcolor{teal}{To analyse divergent thinking performance, experts rated four measures using standard methods of design research~\cite{Wadinambiarachchi2024TheThinking, Shah2003MetricsEffectiveness, Nelson2009RefinedEffectiveness}. These measures included \textsc{fluency}: the number of products created by each participant~\cite{Wadinambiarachchi2024TheThinking, Shah2003MetricsEffectiveness}, \textsc{variety}: coverage of the solution space during ideation~\cite{Nelson2009RefinedEffectiveness},\textsc{originality}: infrequency of an idea relative to other participants' responses~\cite{Nelson2009RefinedEffectiveness}, and \textsc{quality}: the usefulness and feasibility of the product rated by the experts.}
Two expert raters, blind to experimental conditions, evaluated the products. They brought complementary perspectives: one was a university lecturer with 11 years' experience in user-centred design, while the other had 8 years of industry experience in product and service design. Their ratings exhibited almost perfect agreement ($\kappa$ = 0.87). Clustering, ratings and all questionnaire responses were imported to \texttt{R}\footnote{https://www.r-project.org} for statistical analysis.

\subsection{Data Analysis}

\subsubsection{Quantitative Analysis}

Post-task questionnaires and expert ratings were analysed using Bayesian statistical methods~\cite{RichardMcElreath2015StatisticalRethinking, Kay2016Researcher-CenteredHCI}. Unlike traditional significance testing, this approach allows for a nuanced interpretation of differences between conditions by estimating effect probabilities and uncertainty. Its flexibility and capacity for extensibility make it well-suited for the small sample sizes common in HCI research~\cite{Kay2016Researcher-CenteredHCI,Schmettow2021NewResearchers, Li2025EstimatingInteraction}, ensuring statistically stable preliminary findings.

Following established practices~\cite{Wadinambiarachchi2024TheThinking, Brailsford2025ResponsibilitySystems, Li2025EstimatingInteraction}, we implemented Bayesian regression via the \texttt{brms} package~\cite{burker2017brms} in \texttt{R}\footnote{An interface to the Stan probabilistic programming language~\cite{carpenter2017stan}}. We used multivariate models for questionnaire outcomes and multilevel models for single outcomes, including participants as random effects to account for repeated measures. We employed weakly informative regularizing priors, $Normal(0, 2)$, to ensure data-driven inference. Model stability and convergence were confirmed via standard diagnostics: R hat ($\hat{R} < 1.01$)~\cite{vehtari2021rhat} and Effective Sample Size ($ESS > 1000$)~\cite{burker2017brms}. All our estimates fulfilled these criteria.

We report our results using posterior means, standard deviations, and 89\% credible intervals, which differ from frequentist confidence intervals, consistent with McElreath’s~\cite{RichardMcElreath2015StatisticalRethinking} recommendations and recent HCI literature~\cite{Wadinambiarachchi2024TheThinking,Li2025EstimatingInteraction,Brailsford2025ResponsibilitySystems}. For hypothesis testing, we calculated Bayes factors\footnote{Also known as Evidence Ratios.} \textcolor{teal}{and report them on the natural log scale ($\ln(BF)$). All hypotheses are framed directionally (e.g., \texttt{Sketch} > \texttt{Text}), based on our expectation that sketching would outperform the text baseline. Positive $\ln(BF)$ values indicate evidence in favour of the hypothesised direction (e.g., \texttt{Sketch} > \texttt{Text}), negative values indicate evidence in favour of the opposing direction (e.g., \texttt{Sketch} < \texttt{Text}), and the magnitude reflects the strength of evidence in both cases. We interpret the evidence strength using Table~\ref{tab:bf}, which provides $\ln(BF)$ thresholds adapted from \citet{wagenmakers2011psychologists}.} For clarity, we plotted Posterior Estimated Means and Highest Posterior Density (HPD) intervals back-transformed to the original scale of each variable (e.g., the 1--7 CSI scale). Full modelling reports are provided in the supplemental material. Note that p-values are not used in Bayesian statistics, and no claims regarding ``statistical significance'' should be derived from our results.

\subsubsection{Qualitative Analysis}

We analysed qualitative data (interview-transcripts, sketches and AI-generated images) using Braun and Clarke's six-phase reflexive thematic analysis approach (RTA)~\cite{Terry2021EssentialsAnalysis, Braun2022ThematicGuide}, in constructing themes (meaning-based patterns) to report our interpretations of the data~\cite{Terry2021EssentialsAnalysis, Braun2022ThematicGuide}. \textcolor{teal}{Audio recordings were automatically transcribed using Microsoft Word and verified and corrected by the first author. The first author then coded the interview data with an inductive, data-driven approach where codes operated at both semantic and latent levels (semantic codes captured surface-level meanings close to participants' language, while latent codes captured implicit meanings). Participant sketches, sketches and tags as well as corresponding AI-generated images were also used to support the coding process. These codes were grouped to develop 5 candidate themes: \textit{AI for explorations and for validation: Emerging behaviours}, \textit{Different interfaces for different design thinking stages}, \textit{Inspiring for some, but frustrating for others: contradicting statements about sketch-based interfaces},\textit{ Preconceived expectations for what AI outcomes}, and \textit{Participants' expectations for future idea explorations}. These themes were then refined through iterative discussions among all authors, resolving overlaps, removing redundancies, and interpreting insights beyond the surface-level meaning. These were further clustered into 3 overarching themes that synthesised and extended the initial themes. As RTA requires researchers to reflect on how their own experiences shape the analytical process~\cite{Braun2022ThematicGuide}, we have included our positionality statement in Appendix~\ref{sec:position}.}

\section{Findings}

We present our results in two parts: descriptive trends from the Bayesian statistical analysis and qualitative insights from Reflexive Thematic Analysis. 

\begin{figure}[tb]
    \centering
    \includegraphics[width=.85\linewidth]{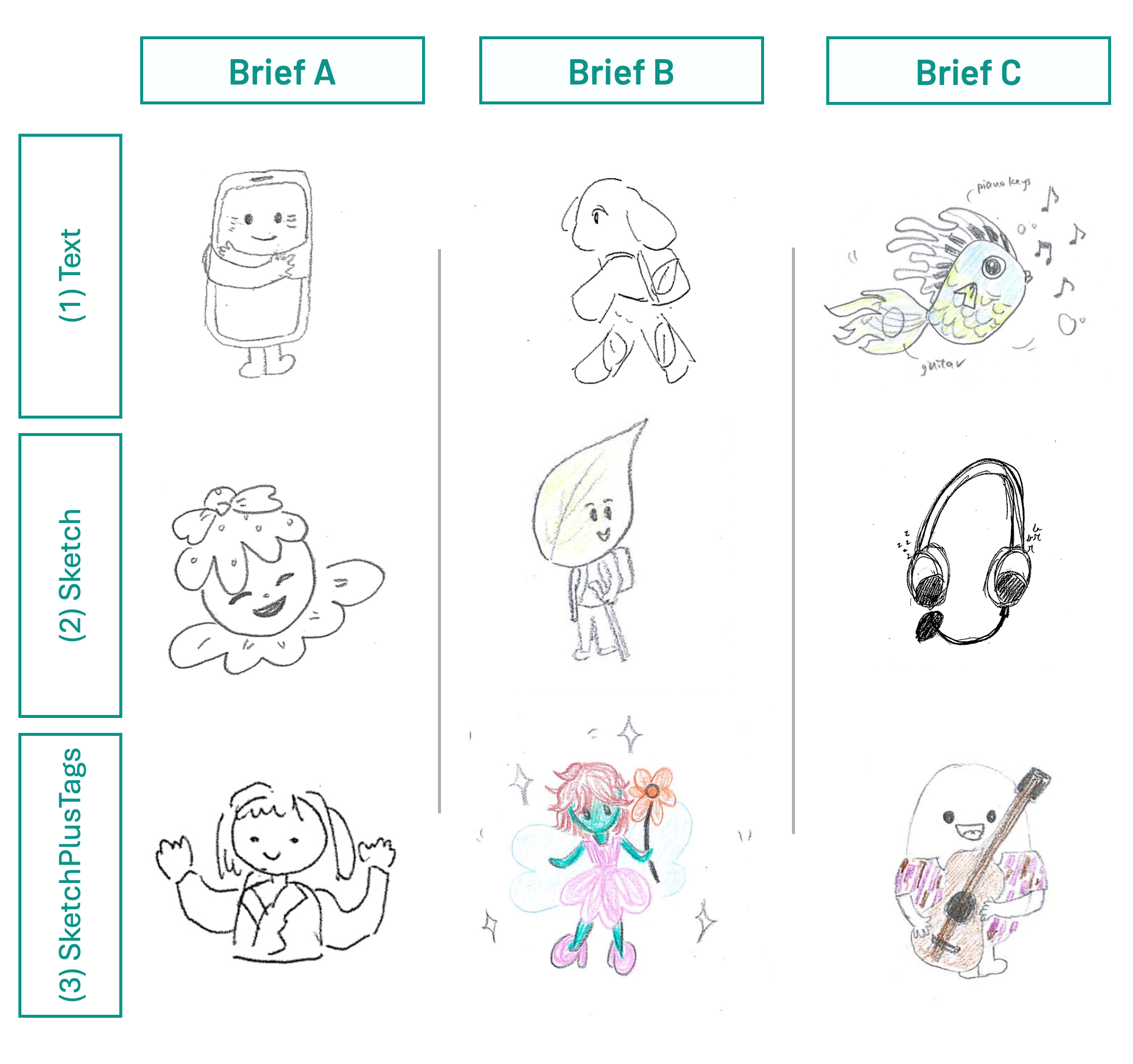}
    \caption{Example products created by participants in each condition \textcolor{teal}{for each brief (columns: Brief A, Brief B, Brief C) and input modality (rows: (1) \texttt{Text}, (2) \texttt{Sketch}, (3) \texttt{SketchPlusTags}).}}
    \Description{This figure shows a 3×3 grid of mascot character examples produced by participants, organised by brief (columns) and input modality (rows). Rows correspond to input modality — Text, Sketch, and SketchPlusTags — and columns correspond to brief topic. For Brief A (social connectivity): in the Text condition, a participant produced a smartphone-shaped character with a face and arms; in the Sketch condition, a participant produced a flower-like character with petals, a bow, and a smiling face; and in the SketchPlusTags condition, a participant produced a girl character with large ears, waving hands, and a jacket. For Brief B (outdoor/nature): in the Text condition, a participant produced a dog-like walking character; in the Sketch condition, a participant produced a leaf-shaped character with a face wearing clothes; and in the SketchPlusTags condition, a participant produced a coloured fairy character with pink hair holding a flower, surrounded by sparkles. For Brief C (music): in the Text condition, a participant produced a fish-like character surrounded by musical notes, annotated with the words ``piano keys'' and ``guitar''; in the Sketch condition, a participant produced a headphones character with a face; and in the SketchPlusTags condition, a participant produced a round ghost-like character playing a guitar. }
    \label{fig:participantssketches}
\end{figure}

Overall, participants created a \textbf{total of 177 products, 55 in the \texttt{Text} condition, 68 in the \texttt{Sketch} condition, and 54 in the \texttt{SketchPlusTags} condition} (see Figure~\ref{fig:participantssketches}). \textcolor{teal}{The variation in number of products across conditions was an emergent outcome of the ideation activity, which we explored further in the \textsc{fluency} analysis (see Section~\ref{sec:Fluency}).}

\subsection{Expressing Intent}

To address RQ1, we operationalised \textsc{perceived support for expressing Intent} as the extent to which participants felt the input modality allowed them to define the direction and scope of their intent to \texttt{SketchifAI}. We modelled responses to both questions in the same Bayesian multivariate cumulative-probit model (see Table~\ref{tab:clarity_model}).

\begin{table*}[t]
    \centering
    \caption{Summary of the cumulative-probit multivariate model for perceived support in clearly expressing intent to AI systems: \texttt{Defining the direction $\sim$ Condition + (1 + Condition | id)} and \texttt{Defining the scope $\sim$ Condition + (1 + Condition | id)}. We report the posterior means ($M$), standard deviations ($SD$) of parameter estimates on the \textit{latent scale,} and 89\% credible intervals (CIs). Unlike frequentist confidence intervals, these represent the 89\% central percentile of the posterior distribution. For brevity, threshold (cut-point) estimates are omitted but provided in the supplementary material. All parameters demonstrated strong convergence, with R-hat = 1.00 and effective sample sizes ESS exceeding 1000.}
    \begin{tabular}{lrrrrr}
        \toprule
        & \multicolumn{2}{c}{\textbf{Sketch}} & \multicolumn{2}{c}{\textbf{SketchPlusTags}} \\
        \cmidrule(lr){2-3} \cmidrule(lr){4-5}
        & \textbf{Est.\ (SD)} & \textbf{89\% CI} & \textbf{Est.\ (SD)} & \textbf{89\% CI} \\
        \midrule
        \textbf{Defining the direction} & $\mathbf{-1.44\ (1.08)}$ & $\mathbf{[-3.35,\ -0.03]}$ & $-0.30\ (0.95)$ & $[-1.78,\ 1.13]$ \\
        \textbf{}\textbf{Defining the scope}     & $\mathbf{-2.02\ (1.19)}$ & $\mathbf{[-4.11,\ -0.43]}$ & $-0.26\ (0.73)$ & $[-1.41,\ 0.83]$ \\
        \bottomrule
    \end{tabular}
    \label{tab:clarity_model}
\end{table*}

\begin{figure}[tb]
    \centering
    \includegraphics[width=.95\linewidth]{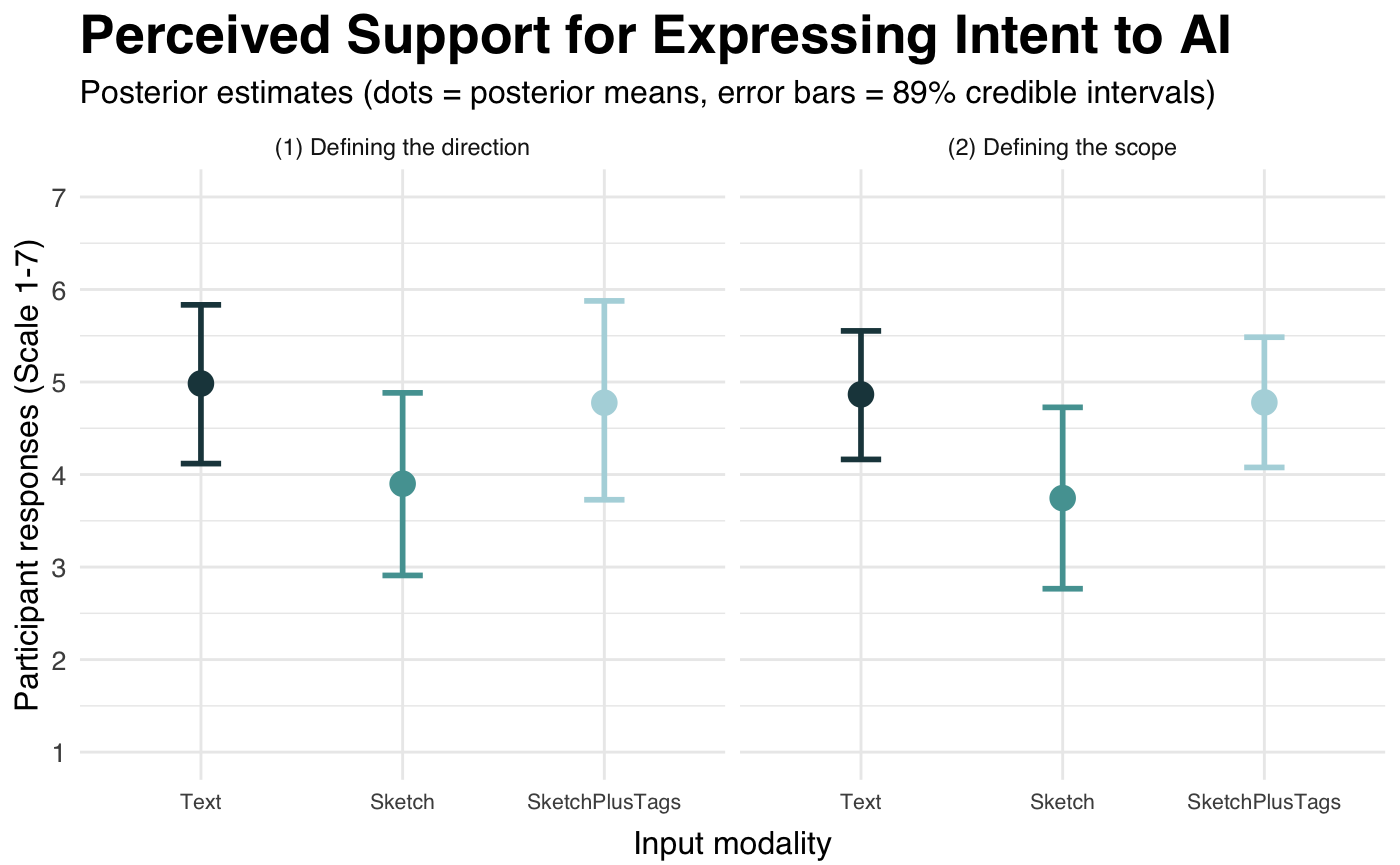}
    \caption{Model posterior predictions for perceived support for \textsc{Expressing intent} clearly to AI. \textcolor{teal}{ (1) \texttt{Defining the direction} (left), (2) \texttt{Defining the scope} (right)}. Error bars represent the standard error of the estimates.}
    \Description{This plot shows model posterior predictions for perceived support for expressing intent clearly to AI, across three input modalities — Text, Sketch, and SketchPlusTags. Results are shown for two sub-dimensions: Defining the Direction and Defining the Scope. Each dot represents a posterior mean and error bars represent 89\% credible intervals.}
    \label{fig:Clarity_expression}
\end{figure}

Our model indicated that \texttt{Sketch} condition had a 95\% probability of yielding lower ratings compared to \texttt{Text} when defining direction ($M=-1.44$, 89\%CI [-3.35, -.03]), \textcolor{teal}{($ln(BF)=-3.00$, $BF=0.05$), strong evidence suggesting} participants found it credibly harder to express their intended direction via \texttt{Sketch}ing alone. 

For defining scope, the deficit of sketching was even more pronounced. The model suggested a 98\% probability that \texttt{Sketch} led to lower ratings relative to baseline ($M=-2.02$, 89\%CI [-4.11, -0.43]) and \textcolor{teal}{($ln(BF)=-3.91$,$BF=0.02$) provided very strong evidence that sketching alone restricted expression of scope compared to the \texttt{text} baseline}. However, we observed \texttt{SketchPlusTags} condition did not differ reliably from \texttt{Text} baseline ($M=-0.26$, 89\%CI [-1.41,  0.83]). Notably, \textcolor{teal}{($ln(BF)= 3.24$,$BF=25.49$)} provided strong evidence that there's a 96\% probability that \texttt{SketchPlusTags} reliably outperformed Sketch alone ($M=1.76$, 89\%CI [0.52, 3.14]).~\textbf{In summary, our findings suggest that sketching alone created a credible hurdle for expressing intent, but this limitation was effectively mitigated by adding tags}.

\subsection{Creativity Support}

\begin{table*}[t]
    \centering
    \caption{Summary of the Multivariate cumulative-probit models for six CSI subscales (\textit{Collaboration, Enjoyment, Exploration, Expressiveness, Immersion, Results-Worth-the-effort}) with \texttt{score $\sim$ Condition + (1 + Condition | id)}. For brevity, threshold (cut-point) estimates are omitted but provided in the supplementary material. We report the posterior means ($M$), standard deviations ($SD$) of parameter estimates on the \textit{latent scale,} and 89\% credible intervals (CIs). Unlike frequentist confidence intervals, these represent the 89\% central percentile of the posterior distribution. All parameters demonstrated strong convergence, with R-hat = 1.00 and effective sample sizes ESS exceeding 1000.}
    \begin{tabular}{lrrrrr}
        \toprule
        & \multicolumn{2}{c}{\textbf{Sketch}} & \multicolumn{2}{c}{\textbf{SketchPlusTags}} \\
        \cmidrule(lr){2-3} \cmidrule(lr){4-5}
        & \textbf{Est.\ (SD)} & \textbf{89\% CI} & \textbf{Est.\ (SD)} & \textbf{89\% CI} \\
        \midrule
        Collaboration:         & $-0.59\ (1.37)$ & $[-2.78,\ 1.54]$ & $-0.15\ (1.47)$ & $[-2.45,\ 2.20]$ \\
        \textbf{Enjoyment:}           & $\mathbf{-2.97\ (1.43)}$ & $\mathbf{[-5.51,\ -1.03]}$ & $0.39\ (0.79)$ & $[-1.61,\ 0.82]$ \\
        \textbf{Exploration:}          & $\mathbf{-2.40\ (1.25)}$ & $\mathbf{[-4.49,\ -0.67]}$ & $1.05\ (0.85)$ & $[-2.44,\ 0.16]$ \\
        Expressiveness:        & $-1.12\ (1.11)$ & $[-2.96,\ 0.44]$ & $0.34\ (0.81)$ & $[-0.85,\ 1.64]$ \\
        Immersion:             & $0.67\ (1.10)$ & $[-0.91,\ 2.50]$ & $-1.24\ (1.44)$ & $[-3.64,\ 0.82]$ \\
        Results Worth the Effort & $0.69\ (1.03)$ & $[-0.79,\ 2.41]$ & $0.40\ (1.39)$ & $[-1.66,\ 2.60]$ \\
        \bottomrule
    \end{tabular}
    \label{tab:CSI_model}
\end{table*}

\begin{figure}[tb]
    \centering
    \includegraphics[width=.95\linewidth]{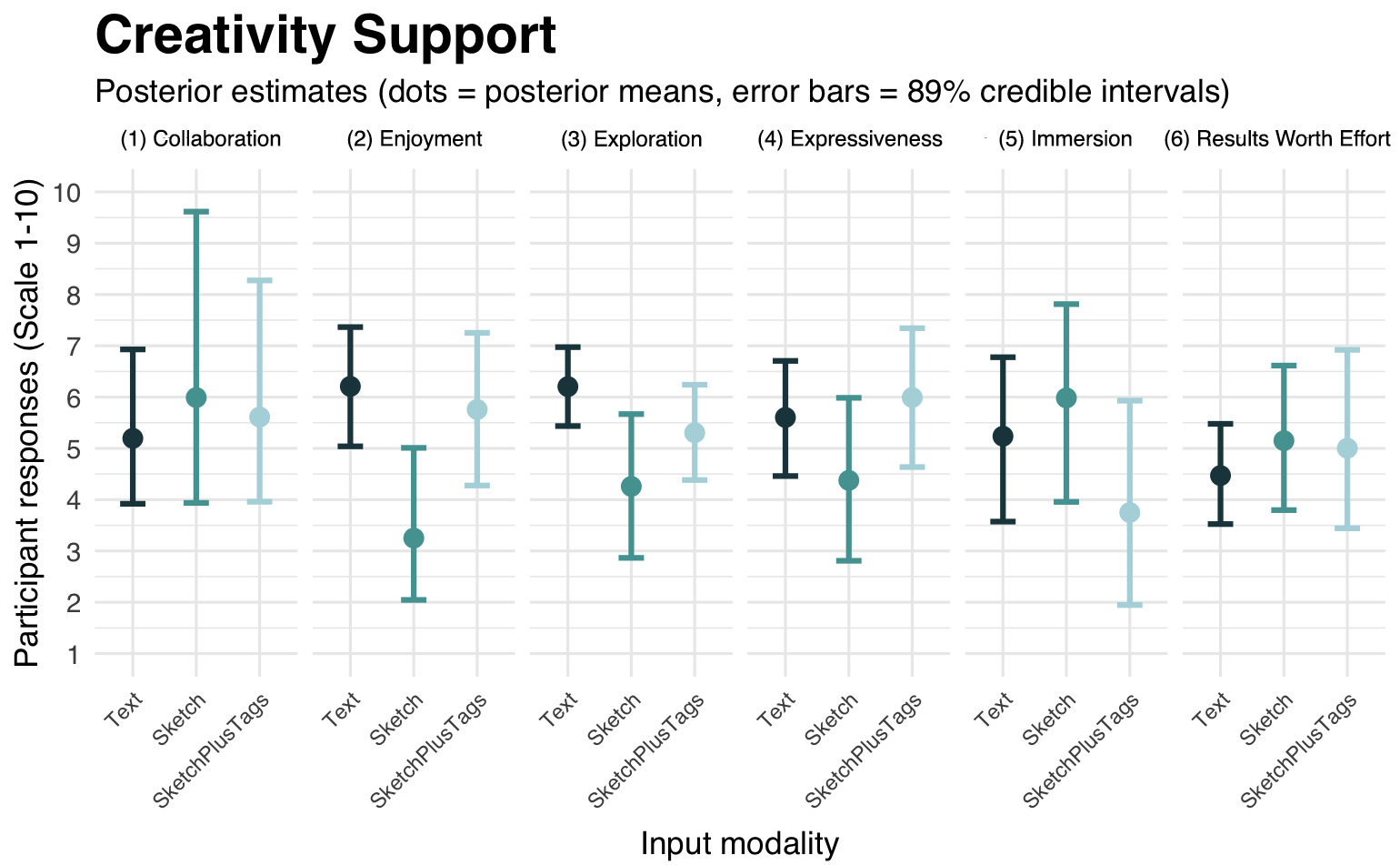}
    \caption{{Model posterior predictions for \textsc{Creativity Support}. \textcolor{teal}{ (1) \texttt{Collaboration}, (2) \texttt{Enjoyment}, (3) \texttt{Exploration}, (4) \texttt{Expressiveness}, (5) \texttt{Immersion}, (6) \texttt{Results-worth-the-Effort}}. Error bars represent the standard error of the estimates.}}
    \Description{This plot shows model posterior predictions for Creativity Support across three input modalities — Text, Sketch, and SketchPlusTags. Results are shown for six subscales: Collaboration, Enjoyment, Exploration, Expressiveness, Immersion, and Results-Worth-the-Effort. Each dot represents a posterior mean and error bars represent 89\% credible intervals.}
    \label{fig:CSI_plot}
\end{figure}

To address RQ2, we utilised the \textsc{Creativity Support Index} (CSI)~\cite{CherryQuantifyingIndex}. This instrument evaluates how effectively a tool supports the creative process across six-dimensions: Collaboration, Exploration, Expressiveness, Immersion, Enjoyment, and Results Worth Effort. We modelled these subscales using a Bayesian multivariate cumulative-probit model to compare the impact of input modality (see Table~\ref{tab:CSI_model}). 

Our findings revealed that the \texttt{Sketch} condition resulted in reliably lower ratings for Enjoyment ($M=-2.97$, 89\%CI [-5.51, -1.03]) and Exploration ($M=-2.40$, 89\%CI [-4.49, -0.67]) compared to the \texttt{Text} baseline. Our model indicated a 99\% posterior probability that sketching alone reduces enjoyment, \textcolor{teal}{($ln(BF)=-4.61$, $BF=0.01$)} suggesting extreme evidence for this effect. Similarly, a 99\% probability and \textcolor{teal}{($ln(BF)=-4.51$, $BF=0.011$)} provided very strong evidence that sketching hindered exploration. The \texttt{SketchPlusTags} condition mitigated these negative effects to varying degrees. \textcolor{teal}{For the Enjoyment subscale, \textcolor{teal}{($ln(BF)=-0.89$, $BF=0.41$)} indicated only anecdotal evidence for a difference from the \texttt{Text} baseline}. For Exploration, \textcolor{teal}{ moderate evidence indicated that a deficit remained for \texttt{SketchPlusTags} relative to \texttt{Text}} ($M= -1.05$, 89\%CI [-2.02, -0.11]) \textcolor{teal}{($ln(BF)=-2.40$, $BF=0.091$)}. A direct comparison between the two sketching modalities confirmed that \texttt{SketchPlusTags} provided a reliable improvement over pure sketching ($M=1.35$ , 89\%CI [0.01, 2.83]), with a \textcolor{teal}{($ln(BF)=2.1$, $BF=8.211$)} providing moderate support for this recovery.

For all other subscales -- Collaboration, Expressiveness, Immersion, and Results Worth Effort -- the credible intervals included zero, indicating that these qualities were perceived comparably across modalities. Thus, we cannot reliably claim that Sketching offers a distinct advantage or disadvantage for these subscales of creativity support compared to \texttt{Text}. \textbf{In summary, pure sketching created a distinct shortfall in enjoyment and exploration scores. This suggests that without proper semantic grounding, the ambiguity of sketching introduces friction that slows down participants arriving at an AI-generated solution. However, by anchoring sketches with tags, \textcolor{teal}{the \texttt{SketchPlusTags} modality restored user enjoyment to levels comparable to the \texttt{Text} baseline, though a moderate deficit in exploration remained, suggesting that tags partially mitigated the frustration that was inherent in sketching alone.}}

\subsection{Divergent Thinking Performance}

To answer RQ3, we calculated the \textsc{fluency}, \textsc{Variety}, \textsc{Originality}, and \textsc{Quality} of the products based on expert ratings to examine how inspiration stimuli generated through different modalities affect participants' divergent thinking performance (see Table~\ref{tab:DTALL}).

\begin{table*}[tb]
    \centering
    \caption{Summary of Bayesian model results for Fluency, Variety, Originality, and Quality. Fluency was analysed using a negative binomial model; Variety, Originality, and Quality were analysed using Gaussian or linear models as appropriate for the scale. All models included random intercepts and slopes by participant (1 + Condition | id) to account for individual variability in response to conditions following model structure: \texttt{Metric $\sim$ Condition + (1 + Condition | id)}. We report posterior means ($M$), standard deviations ($SD$), and 89\% credible intervals (CIs). These CIs represent the 89\% central percentile of the posterior distribution, differing from frequentist confidence intervals by providing a direct probability of the parameter's location. All parameter estimates demonstrated strong convergence, with Rhat = 1.00 and effective sample sizes (ESS) exceeding 1000.
    \textit{*Note:For the cumulative Quality model, threshold estimates are reported in the supplementary material-C.}}
    \begin{tabular}{llcccccc}
        \toprule
        \multicolumn{2}{c}{} & \multicolumn{2}{c}{\textbf{Intercept}} & \multicolumn{2}{c}{\textbf{Sketch}} & \multicolumn{2}{c}{\textbf{SketchPlusTags}} \\
        \cmidrule(lr){3-4}\cmidrule(lr){5-6}\cmidrule(lr){7-8}
        \textbf{Metrics} & \textbf{Model Family}  & \textbf{Est.\ (SD)} & \textbf{89\% CI}  & \textbf{Est.\ (SD)} & \textbf{89\%CI} & \textbf{Est.\ (SD)} & \textbf{89\% CI} \\
        \midrule
        \textbf{Fluency }    & Neg.\ Binomial & 1.80\,(0.23) & {[}1.42, 2.17{]} & 0.20\,(0.33) & {[}-0.33, 0.72{]} & -0.04\,(0.33) & {[}-0.57, 0.49{]} \\
        Variety     & Gaussian       & 0.22\,(0.04) & {[}0.15, 0.28{]} & -0.02\,(0.07) & {[}-0.13, 0.08{]} & 0.02\,(0.09) & {[}-0.12, 0.16{]} \\
        \textbf{Originality} & Gaussian       & 0.79\,(0.03) & {[}0.74, 0.83{]} & -0.02\,(0.04) & {[}-0.08, 0.04{]} & \textbf{-0.06\,(0.04)} & \textbf{{[}-0.12, -0.00{]}} \\
        Quality & Cumulative       & -* & -* & -0.24 (0.50)  & {[}-1.06, 0.52{]} & 0.14\,(0.37) & {[}-0.44, 0.73{]} \\
        \bottomrule
    \end{tabular}
        \label{tab:DTALL}
\end{table*}

\label{sec:Fluency}
\begin{figure}[tb]
\centering
    \includegraphics[width=.435\textwidth]{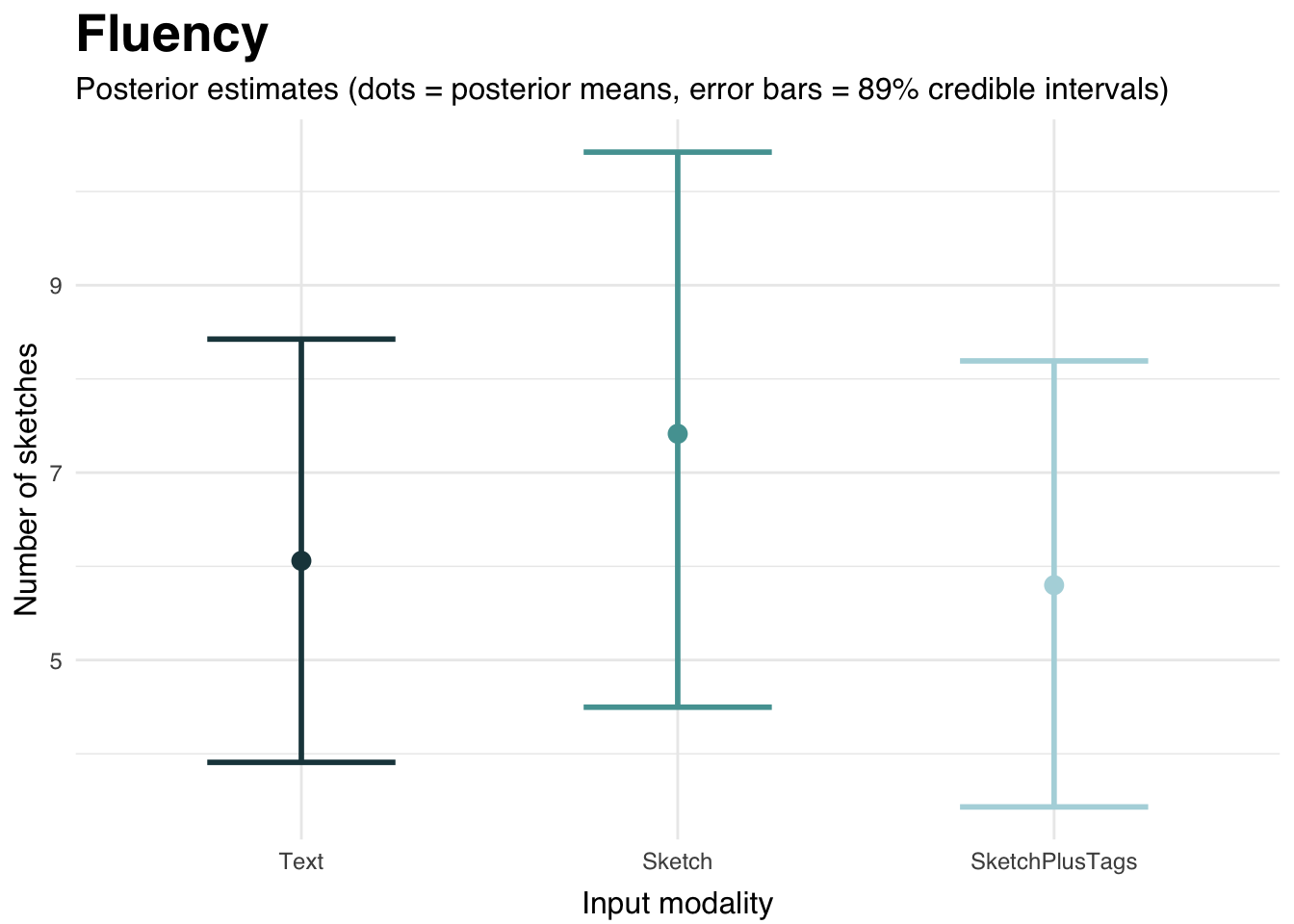}
    \caption{Model posterior predictions for \textsc{Fluency} (Number of products generated). Error bars represent the standard error of the estimates.}
    \Description{This plot shows model posterior predictions for Fluency, operationalised as the number of products generated, across three input modalities — Text, Sketch, and SketchPlusTags. Each dot represents a posterior mean and error bars represent 89\% credible intervals}
    \label{fig:fluency_plot}
\end{figure}

\subsubsection{Fluency}

To model \textsc{Fluency}, we considered the number of products created by each participant~\cite{Wadinambiarachchi2024TheThinking, Shah2003MetricsEffectiveness}. We used a negative binomial distribution with a log link. This approach allows us to estimate the extent to which input modality influences the mean frequency of products produced. The Bayesian model estimated \textsc{Fluency} of the \texttt{Text} condition at mean = 1.80 on the log scale (89\%CI [1.43,2.19]). Relative to this baseline, \texttt{Sketch} ($M=0.21$, 89\%CI [-0.33, 0.75]) and \texttt{SketchPlusTags} ($M=-0.04$, 89\%CI [-0.56, 0.49]) had wide credible intervals overlapping zero, indicating little evidence for differences. Complementing this, the hypothesis tests suggested a 74\% probability that \texttt{Sketch} has a positive effect relative to \texttt{Text} \textcolor{teal}{($ln(BF)= 1.03$, $BF=2.80$)}. In contrast, \texttt{SketchPlusTags} had only a 45\% probability of being greater than zero, \textcolor{teal}{($ln(BF)= -0.20$, $BF=0.82$)}. \textbf{Taken together, the \texttt{Sketch} condition shows a modest tendency toward higher fluency, although the posterior predictions and hypothesis tests indicated substantial uncertainty, with no reliable differences between other conditions.}

\subsubsection{Variety}
To compute the variety score we used the following method~\cite{Wadinambiarachchi2024TheThinking} that uses expert ratings:

 {\small
    \begin{equation}
    \begin{split}
    \text{Variety} = \frac{\text{Number of clusters that a participant's products belong to - 1}}{\text{Number of clusters - 1}}\\
    \end{split}
    \label{eq:variety}
    \end{equation}
    }

\begin{figure}[b]
\centering
    \includegraphics[width=.435\textwidth]{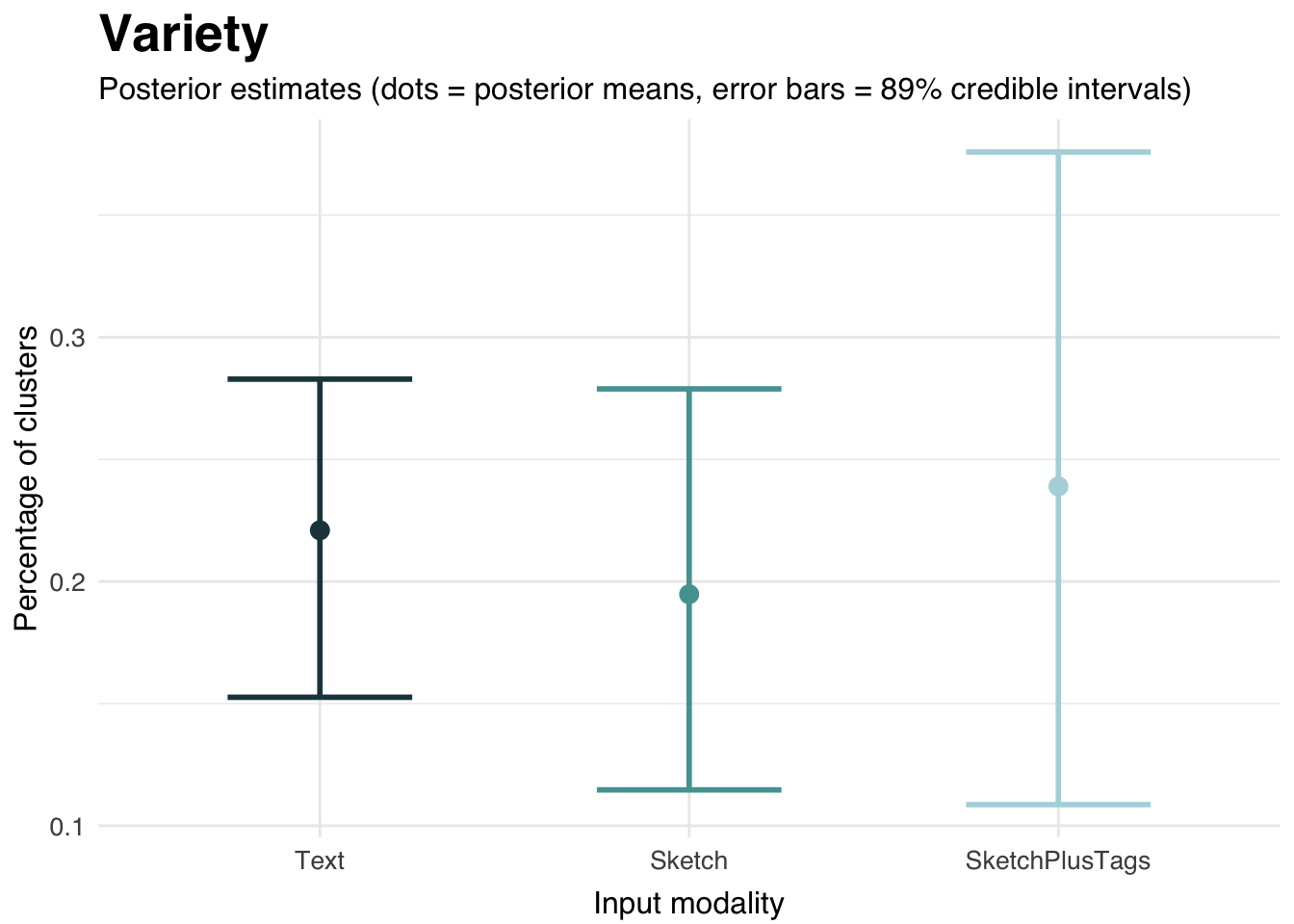}
    \caption{Model posterior predictions for \textsc{Variety} (Number of clustered covered). Error bars represent the standard error of the estimates.}
    \Description{This plot shows model posterior predictions for Variety, operationalised as the number of concept clusters covered, across three input modalities — Text, Sketch, and SketchPlusTags. Each dot represents a posterior mean and error bars represent 89\% credible intervals.}
    \label{fig:variety_plot}
\end{figure}

We computed \textsc{Variety} using the variety score (\ref{eq:variety}). We limited our analysis to simpler models, emphasising the estimation of direct relationships instead of pursuing full causal mediation due to our sample size constraints. We employed a Bayesian linear regression model.

We modelled \textsc{Variety} using a Gaussian family with varying intercepts and slopes by participant. The group-level standard deviations indicate modest variation across participants, with greater variability for the \texttt{SketchPlusTags} condition ($M=-0.02$, 89\%CI [ -0.12, 0.15]) compared to \texttt{Sketch} ($M=-0.02$, 89\%CI [ -0.13, 0.08]) and the intercept ($M= 0.22$, 89\%CI [ 0.15, 0.28]). \textcolor{teal}{Posterior probabilities indicated a 34\% likelihood that \texttt{Sketch} outperforms \texttt{Text} ($ln(BF)= -0.65$,  $BF=0.52$, anecdotal evidence) and a 58\% likelihood that \texttt{SketchPlusTags} outperforms \texttt{Text}, ($ln(BF)= 0.34$, $BF=1.41$, anecdotal evidence), indicating neither result is sufficient to draw a firm conclusion in either direction. \textbf{Overall, the data provide inconclusive evidence as to whether the modality reliably alters the breadth of the products across conditions.}}

\subsubsection{Originality}
We calculated originality scores using the methods suggested in prior literature on divergent thinking evaluation~\cite{Wadinambiarachchi2024TheThinking, Jansson1991DesignFixation, Guilford1956TheIntellect.} The originality of an output depends on how many other participants created outputs in the same cluster.

  {\small
    \begin{equation}
    \begin{split}
    \text{Originality} = 1-\frac{\text{Number of other participants with products in the cluster}}{\text{Number of other participants}}\\
    \end{split}
    \end{equation}
    \label{eq:originality}
    }

To model \textsc{Originality}, we used expert ratings. However, we restricted the analysis to simpler models, focusing on estimating the direct relationships rather than attempting full causal mediation, because of our sample size. We used a Bayesian linear regression model.

\begin{figure}[tb]
\centering
    \includegraphics[width=.435\textwidth]{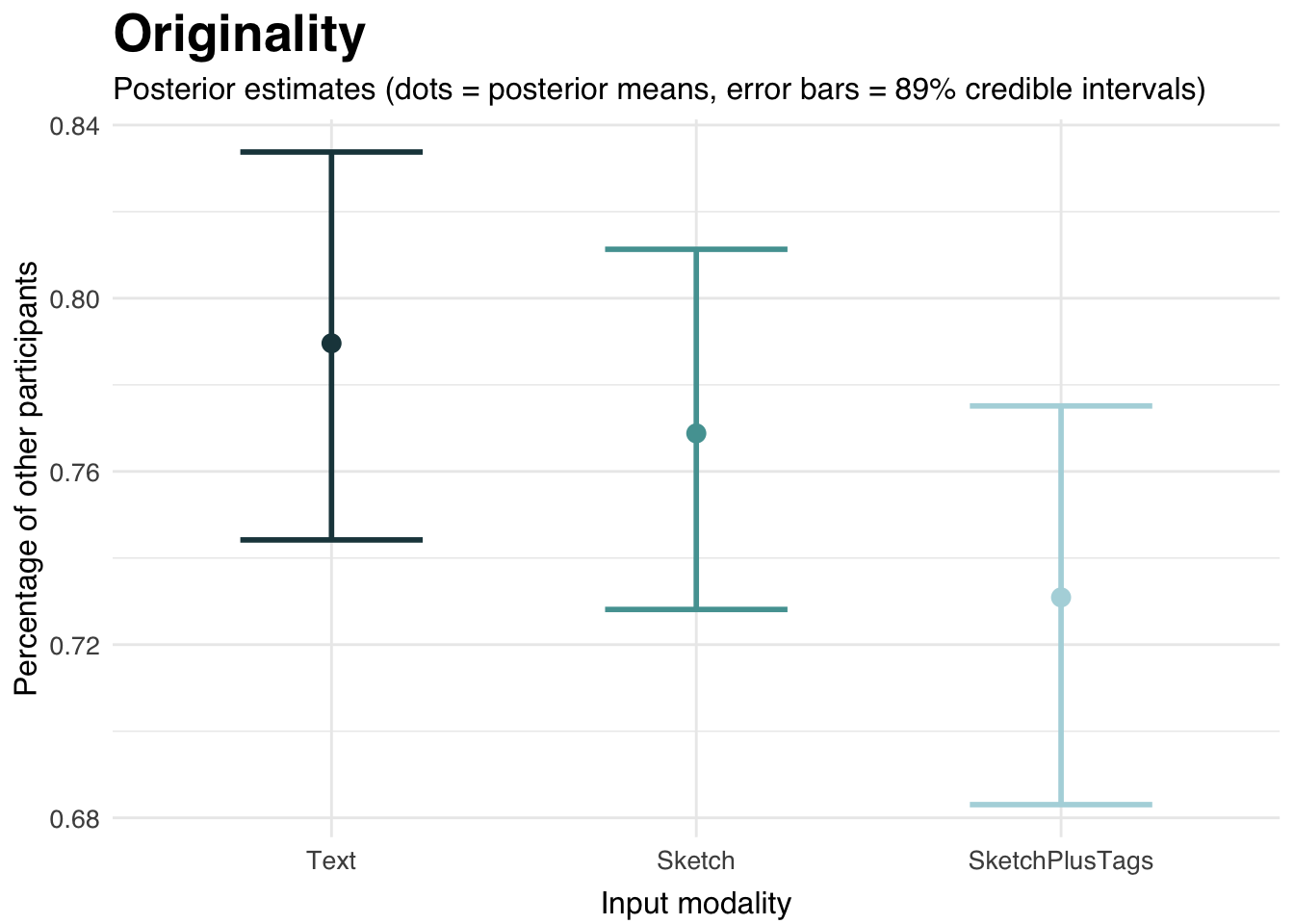}
    \caption{Model posterior predictions for \textsc{Originality} (how unique products are). Error bars represent the standard error of the estimates.}
    \Description{This plot shows model posterior predictions for Originality, operationalised as how unique participant-generated products are, across three input modalities — Text, Sketch, and SketchPlusTags. Each dot represents a posterior mean and error bars represent 89\% credible intervals.}
    \label{fig:originality_plot}
\end{figure}

We modelled the \textsc{Originality} scores with varying intercepts by participant. The average score in the \texttt{Text} condition was estimated at ($M=0.79$, 89\%CI [ 0.74, 0.83]). Relative to this baseline, \texttt{Sketch} was almost unchanged, ($M=-0.02$, 89\%CI [ -0.08, 0.04]), while \texttt{SketchPlusTags} had a slightly lower effect ($M=-0.06$, 89\%CI [ -0.12, 0.00]), with posterior probability of 72\% \textcolor{teal}{ that \texttt{Sketch} reduces originality relative to \texttt{Text}($ln(BF)=-0.94$, $BF=0.39$) suggesting only anecdotal evidence. In contrast, the \texttt{SketchPlusTags} condition exhibited a credible negative effect on originality relative to \texttt{Text}, with a posterior probability of 94\%, ($ln(BF)=-2.81$, $BF=0.06$, moderate evidence). \textbf{In summary, there was insufficient evidence that sketching alone altered originality, while the addition of semantic tags appeared to create a bottleneck that limited the originality of the generated products.}}

\subsubsection{Quality of the Products}
\begin{figure}[tb]
\centering
    \includegraphics[width=.435\textwidth]{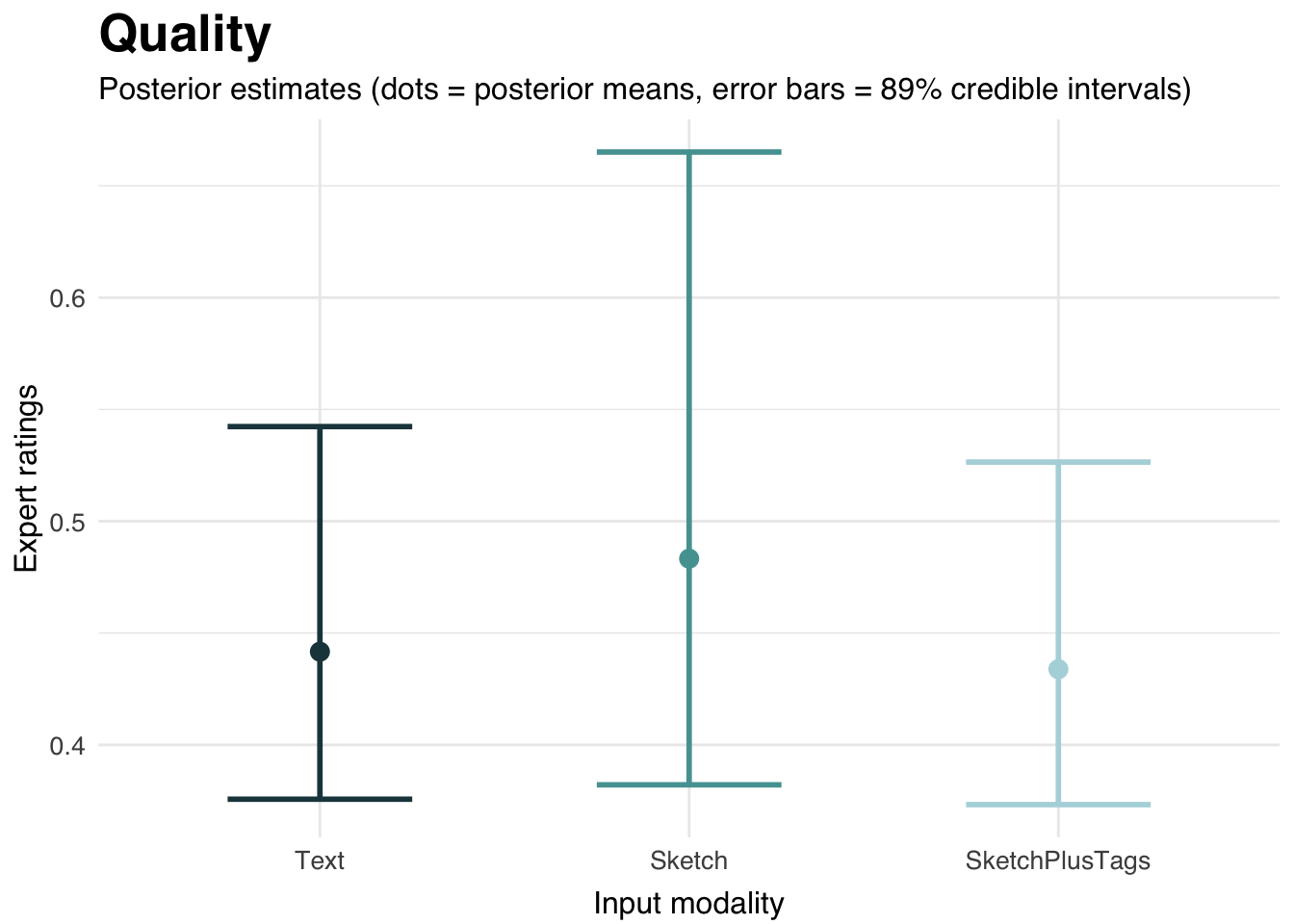}
    \caption{Model posterior predictions for \textsc{Quality} (Expert evaluations). Error bars represent the standard error of the estimates.}
    \Description{This plot shows model posterior predictions for Quality, based on expert evaluations of participant-generated products, across three input modalities — Text, Sketch, and SketchPlusTags. Each dot represents a posterior mean and error bars represent 89\% credible intervals.}
    \label{fig:quality_plot}
\end{figure}

To model \textsc{Quality}, experts evaluated whether the participants' products met the design brief: \texttt{Yes =1, Maybe =0.5, and No =0.} Again, our explorations were limited to simple models because of our sample size restrictions and we opted to explore only the direct effects. The expected value for each response is based on a cumulative probit model. 

In the cumulative probit model, the three intercepts (-0.96, 1.21 and 1.57) mark thresholds on the latent scale that separate the observed response categories. Condition effects describe how responses shift relative to the Text condition. For \texttt{Sketch}, the posterior effect was ($M=-0.24$, 89\%CI [ -1.06, 0.52]). For \texttt{SketchPlusTags}, the effect was ($M=0.14$, (89\%CI [ -0.44, 0.73]), both intervals are wide and include zero. \textcolor{teal}{ Posterior probabilities indicated a 30\% likelihood that \texttt{Sketch} outperforms \texttt{Text}, ($ln(BF)= -0.87$, $BF=0.42$)} providing anecdotal evidence. Similarly, adding tags showed only a 65\% likelihood that \texttt{SketchPlusTags} outperforms \texttt{Text}\textcolor{teal}{($ln(BF)= 0.62$, $BF=1.85$, anecdotal evidence)}. \textbf{In summary, the posterior estimates provided no credible evidence to suggest a reliable shift in quality of products across conditions.}

\subsection{UMUX-Lite}
We modelled UMUX-Lite $\sim$ Condition (1 + Condition |id), in a Bayesian linear regression model. The Text condition was ($M=4.15$ 89\%CI [ 3.56, 4.72]). \texttt{Sketch} produced lower scores than Text ($M=-0.77$, (89\%CI [-1.38, -0.17]), \textcolor{teal}{with 97\% posterior probability that \texttt{Sketch} reduces usability relative to \texttt{Text} ($ln(BF)= -3.50$, $BF=0.03$), strong evidence}. \texttt{SketchPlusTags} was perceived as comparably usable to text ($M=-0.20$, 89\%CI [-0.71, 0.32]),\textcolor{teal}{with 74\% posterior probability that \texttt{SketchPlusTags} reduces usability relative to \texttt{Text} ($ln(BF)= -1.05$, $BF=0.35$), anecdotal evidence}. \textbf{These results suggest that while pure sketching introduces interaction friction, the addition of tags \textcolor{teal}{appeared to restore usability to near-baseline levels, though the evidence remains anecdotal.}}

\subsection{Qualitative Analysis}
To explain the nuances in our statistical findings, such as why participants preferred text over sketch and why the use of sketching \textcolor{teal}{showed a modest tendency towards} improvement in \textsc{Fluency}, but lowered their enjoyment and expressiveness, we conducted a reflective thematic analysis of the interview data, the prompts used (text, sketches, and sketches plus tags) and the AI generated images to paint a holistic picture of this phenomenon. This helped us to identify three core themes: 

\subsubsection{Preconceived Expectations and the Need for Conversation}
Some participants entered the study with expectations that AI should be a state-of-the-art collaborator; consequently, they did not find the prototype helpful for their specific needs. For example, P01 had used a commercial web design product and noted:

\begin{quote}
    \textit{``I’ve tried Lovable\footnote{\url{https://lovable.dev/}} before... what I mean [by] powerful is, like it [AI] can somehow grab your idea and... come up with maybe [a] more optimal solution.''} (P01 | W)
\end{quote}

This response illustrates that some participants were more interested in bypassing the design process to arrive immediately at a production-ready outcome indicating their temptation to arrive at the final solution straight away, rather than engaging in an iterative design journey.

In addition, participants who reported low confidence in sketching found the interface \textit{``frustrating to use''} (P08|w) and not \textit{``[their] cup of tea''} (P09 |w). Others explained that their preference for text over sketches depended on the specific task goal:
\begin{quote}
    \textit{``I tend to use text... [to] branch out into many different aspects... that helps me to broaden my perspective... when I’m first exploring a design''} (P03 |W) .
\end{quote}

This indicate that preference for an interaction modality was shaped by sketching confidence, familiarity with the tool, and the desire to follow path of least resistance, which means choosing the course of action requiring the least cognitive effort to perform a creative task~\cite{Ward1994StructuredGeneration, Cheng2014ADesigners}. 

In addition, several participants mentioned that the lack of conversational ability in \texttt{SketchifAI} made the AI feel like a tool rather than a collaborator. We infer that these factors likely contributed to the low ratings in the Creativity Support Index. Some participants expressed a preference for dialogic engagement where the AI could ask for clarification or offer its own feedback:

\begin{quote}
    \textit{``I feel like if the AI could respond back to me... with its own ideas, that would be helpful as well. It's like a two-way interaction.''} (P09 | W)
\end{quote}

The \texttt{SketchPlusTags} interface emerged as a successful middle ground, providing the semantic grounding that the our prototype otherwise lacked. This was illustrated by P07:
\begin{quote}
    \textit{``Whereas the previous ones were... sort of a gamble... I like the consistency for the sketch plus the text. [It is] the consistency of getting what I wanted, plus extra ideas I didn't even think about''} (P07| S+T-3-B).
\end{quote}

Alternatively, participants suggested audio as a complementary modality. They noted that \textit{``talking through''} their thought process while sketching would help articulate intentions that are difficult to capture via drawing alone.

\begin{quote}
    \textit{``Having an audio input would be a great way... because as you're designing, sometimes to help voice out your ideas, they might not be really fleshed out, but you're just sharing your thinking process with the AI''} (P03| W).
\end{quote}

This highlights participants' expectation for fluid interfaces that allow them to move between thinking aloud their ideas, which will lead to reflection, while making it clearer for the AI to understand their ideas in a natural way.

\subsubsection{AI for Exploration vs. Validation: Emerging Behaviours}
Most participants approached the design task by noting down key ideas from the brief and identifying the overall goal. Once the problem was defined, they followed distinct interaction paths: some drew directly on the digital canvas, while others began with analogue pencil sketches on paper, redrew them with alterations on the digital canvas, and used the AI to extend their initial explorations.

For instance, P03 used the tool to find \textit{``\textit{...relevant objects or ideas for [a] certain topic''}} (P03| S+T21-1-A), while P06 adopted an iterative approach: \textit{``\textit{[I] sketched some elements... and [saw] how AI goes, and then I try to add more elements and see how it works''}} (P06| S-1-A). Other participants focused on semantic keywords from the brief to guide their prompts:

\begin{quote}
    \textit{``My overall thinking process was to focus on the keywords... like eco-friendly, hills, and hiking... the mascot is somewhat based around that idea''} (P08| T-3-B).
\end{quote}

In contrast, some participants sought to use the AI solely for verifying existing ideas--a behaviour that deviates from the intended purpose of a divergent thinking task. These participants sought reassurance that their concepts were ``correct'' or viable by looking at the AI generated image as a simulator of their mental imagery. P08 explained:

\begin{quote}
    \textit{``It [AI] helped me... understand what I'm doing wrong. So, I... [don't] make the same mistake on paper and don't use the paper again and again.''} (P08| S-1-C)
\end{quote}

Similarly, P01 noted: \textit{``I came up with an idea first [and] I drew, and then I kind of validated using [the] system''} (P01| S+T- C)

\begin{figure}[tb]
    \centering
    \includegraphics[width=.75\linewidth]{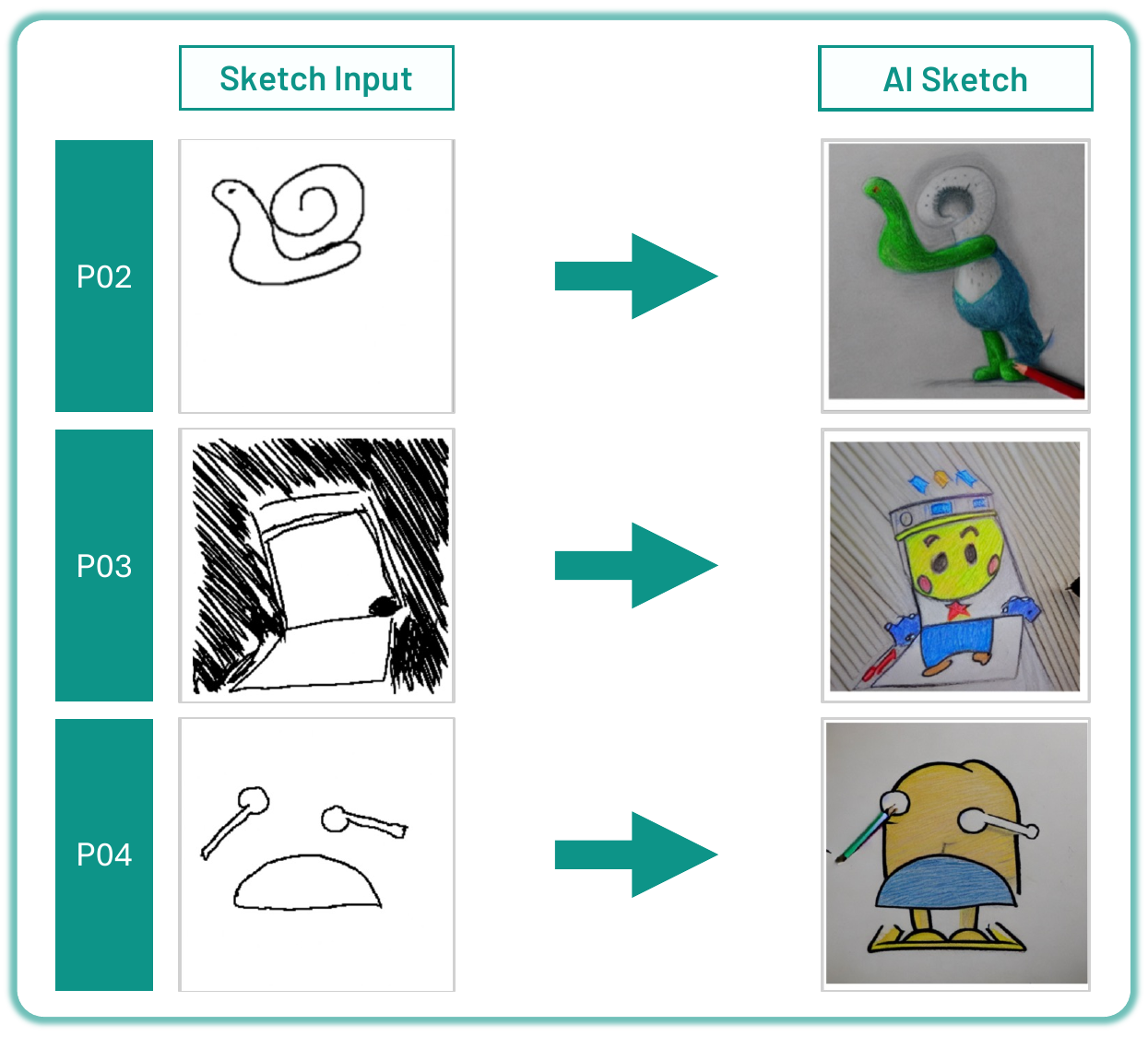}
    \caption{Example sketches input by participants and \textcolor{teal}{corresponding AI-generated sketch outputs} \textcolor{teal}{for participants P02 (top), P03 (middle), and P04 (bottom) in the Sketch condition.}}
    \Description{This figure shows three pairs of sketch inputs and corresponding AI-generated outputs in the Sketch condition, arranged in rows by participant. P02 (top) provided a simple line drawing of a snail, and the AI reinterpreted the spiral shell as a stylised horn and the body as a hybrid of tortoise and bird, producing a bipedal character rather than a snail. P03 (middle) provided a heavily shaded drawing of a pair of beamed musical notes, and the AI generated a coloured cartoon character of a yellow robot-like figure wearing a hat and holding tools. P04 (bottom) provided a minimal line drawing of a drum with drumsticks, and the AI generated a coloured cartoon character of a square-shaped yellow humanoid figure. In all three cases the AI-generated outputs differ substantially from the sketch inputs, illustrating how the system interpreted abstract line drawings as low-to-mid fidelity mascot characters, with the potential to inspire unexpected creative directions.}
    \label{fig:AIMis}
\end{figure}

These behaviours indicate a split in user intent: while some prioritised exploration by asking the AI to generate new symbols and concepts, others utilised the AI for reassurance and visualisation of pre-determined ideas. Some participants were often in a \textit{validation loop}, repeatedly generating images until the AI-output correctly resembled their mental imagery. Because the AI frequently did not deliver the exact idea they wished for based on their sketch outlines, as they tried to keep sketching and generating to get a ``correct'' or ``validated'' result. 

\subsubsection{The Right Tool for the Right Time: Different Interfaces for Different Stages of Design Thinking}

Some participants discussed that distinct interfaces supported different stages of the design process, suggesting there is no single ``silver bullet.'' For example, participants noted that the text modality was best suited to \textit{``help brainstorm''} (P05) and for initial exploration of the problem space, using AI to expand their conceptual solution space. P06 explained:

\begin{quote} 
    \textit{``My favourite [modality] is the one where I just add prompts... I feel like it's very helpful if you're brain blocked... you just type random stuff and it will generate the things that might help you''} (P06| W). 
\end{quote}

For idea expansion, participants preferred the \texttt{SketchPlusTags} interface, as it enabled them to maintain creative agency while allowing the AI to produce serendipitous outcomes to spark ideation.\textcolor{teal}{Here, creative agency refers to the degree of control a designer retains over the ideation process and its outcomes~\cite{Rezwana2023DesigningSystems,Wadinambiarachchi2025ImaginingFutures}.}

\begin{quote}
    \textit{``I think it [\texttt{SketchPlusTags}] allows me to build on my own, brainstorming a little more... I could take more opportunities with the tags, because sometimes... it will generate something unexpected and then I would take that into consideration''} (P06| W).
\end{quote}

\begin{figure}[tb]
    \centering
    \includegraphics[width=.75\linewidth]{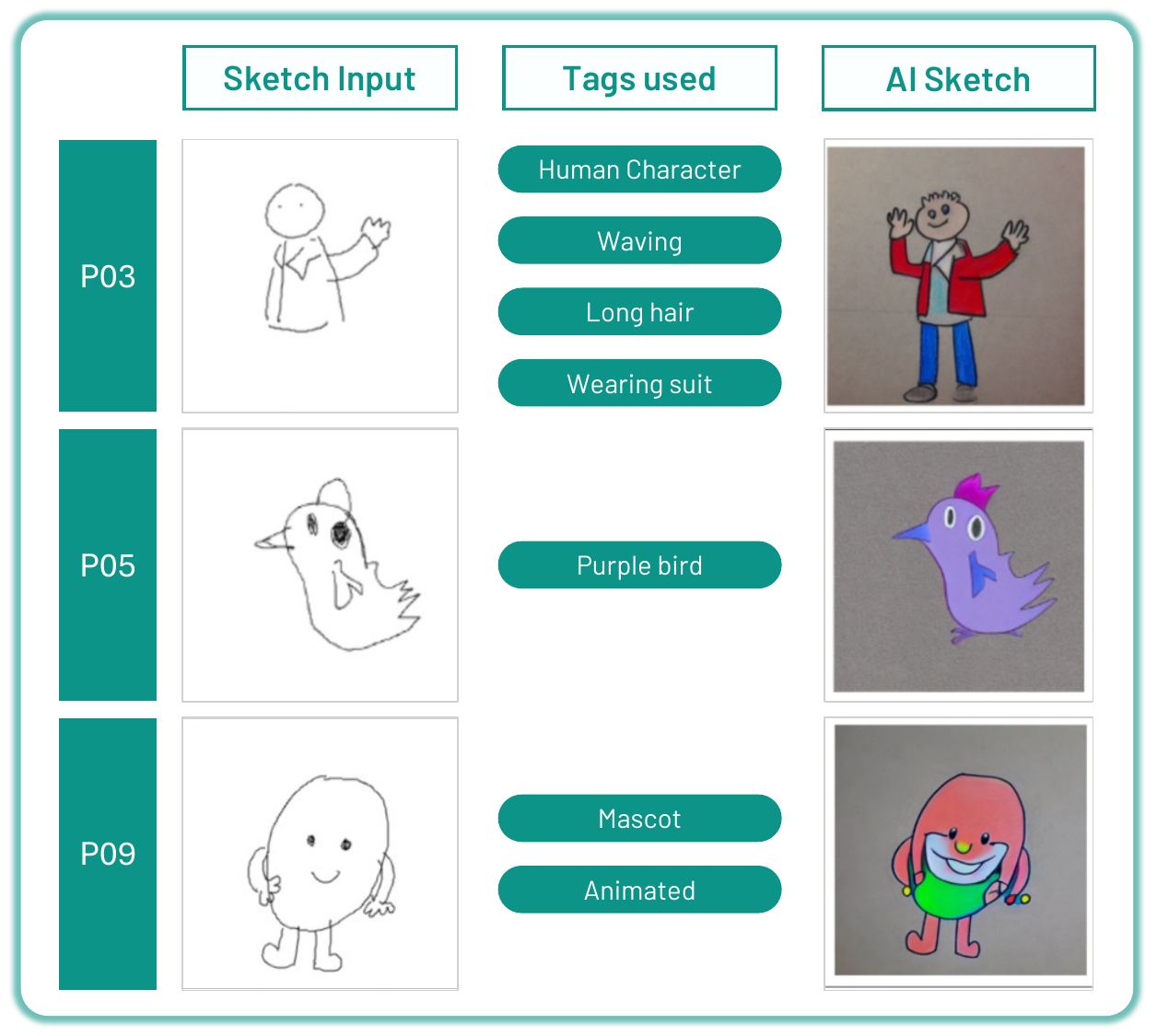}
    \caption{Example sketches inputs, \textcolor{teal}{tags used by participants P03 (top), P05 (middle), and P09 (bottom) and corresponding AI-generated sketch outputs in the \texttt{SketchPlusTags} condition.}}
    \Description{This figure shows three pairs of sketch inputs, tags used, and corresponding AI-generated outputs in the SketchPlusTags condition, arranged in rows by participant. P03 (top) provided a simple line drawing of a waving human figure and used the tags ``Human Character,’’ ``Waving,’’ ``Long hair,’’ and ``Wearing suit,’’ and the AI generated a coloured cartoon character of a person with short hair wearing a red jacket and blue trousers, waving both hands. P05 (middle) provided a line drawing of a bird and used the single tag ``Purple bird,’’ and the AI generated a coloured cartoon of a purple bird with a pink crest. P09 (bottom) provided a simple line drawing of a round smiling figure with small limbs and used the tags ``Mascot’’ and ``Animated,’’ and the AI generated a coloured cartoon character of a round pink and red figure wearing a green outfit. Across all three rows, the tags guided the AI to preserve the general form of the sketch input while adding colour, detail, and character, resulting in outputs that more closely reflect the participant's intent compared to the Sketch condition alone.}
    \label{fig:SketchPlusTags}
\end{figure}

Participants preferred to use the \texttt{Sketch} interface after initial exploration, for refining their ideas, detailed adjustments, visualisation, and concept finalisation. P06 summarised this workflow:

\begin{quote}
    \textit{``It was easier for you to break through blocks when you have prompts, but if you want refinements, it's better to have sketching options''} (P06| W).
\end{quote}

\textcolor{teal}{These responses indicate that participants intuitively adopted different modalities at different stages of the design process. text-prompting was preferred for early divergent exploration, where low effort input allowed rapid idea generation, while sketching was preferred for later refinement stages, where greater commitment to a spatial representation supported more focused, evaluative thinking. The addition of tags in \texttt{SketchPlusTags} appeared to offer a middle ground, supporting idea expansion while maintaining creative agency.}

\section{Discussion}
Through a within participants mixed-methods study, we explored how different input modalities --text-prompting, sketching, and sketching plus tags-- support designers' ability to express their intent to AI tools, their perception of AI's support for creativity, and their divergent thinking performance. \textcolor{teal}{Our quantitative analysis indicated that the \texttt{Sketch} condition tended to enhance ideational fluency, with no credible differences observed in variety, originality, or quality}, compared with the \texttt{Text} baseline, when used to generate inspiration stimuli via \texttt{SketchifAI}. Yet despite the apparent advantages of sketching, students exhibited a strong preference for text-prompting, presenting a paradox for implementing sketch-based interfaces. Our qualitative analysis offers insights into possible causes of these findings. Below, we reflect on the outcomes of this study and discuss implications for design, the limitations of this work, and future directions.

\subsection{Nudging the ``Efficiency Seeker''} 

Most of our participants were undergraduate design students, from an ``AI-savvy'' generation, a demographic accustomed to GenAI tools with text-prompt interfaces. Their experience with publicly available AI tools varied (min = 0.5, max = 2 years; mean = 1.3, SD = 0.56). \textcolor{teal}{Our findings suggest that design students may be becoming less inclined to spend time sketching and experimenting with ideas}, preferring instead to write prompts that drive the AI straight to polished outcomes. This reflects a preference to choose a path of least resistance, a behaviour reinforced by current efficiency-driven AI tools. 

We speculate our findings align with a broader shift: the \textit{Visual Thinker} archetype in design education \textcolor{teal}{seems to be gradually displaced} by an \textit{Efficiency Seeker} mindset that seeks the immediate gratification of high-fidelity text-to-image generation. By offloading the conceptual heavy lifting to GenAI, students may lose the opportunity to exercise the skills vital to building their creative muscle, which, in turn, may lead to cognitive atrophy, diminishing the skills necessary to generate designerly outcomes. In a literature review, ~\citet{laursen2019shortcomings} argue that one methodological approach in designerly thinking is to engage in reflective practices. We \textcolor{teal}{suggest} that the ultimate goal of AI in design education should not be to make the design process easier, but to cultivate a deeper understanding of design thinking and knowledge structures~\cite{oxman1999educating}. 

Therefore, instead of allowing students to use AI that reinforces the temptation to seek instant solutions, we \textcolor{teal}{propose the design of a new genre of AI tools that can serve as scaffolds} for students, by introducing reflective pauses~\cite{cox2016design}. These may prompt students to stop, look, and reflect before solutions are rendered, ensuring the machine's speed does not outpace the student’s development of critical, independent creative judgment. Our suggestions here extend the discourse of slow and reflective AI~\cite{dalsgaard2025thinking,marquardt2025imaginationvellum,van2023objective}. For example, rather than auto-correcting a rough, triangular sketch into a polished image, the AI might highlight the shape and ask the user: \textit{``You've used sharp triangular forms here. In character design, this typically signals 'danger' 'or speed.' Is this intended, or should we explore rounder shapes to convey friendliness?''} \textcolor{teal}{This could potentially move the student from a \textit{Curator Mindset}}, where they merely select from a menu of AI-generated possibilities, to a \textit{Creator Mindset}, where they actively design. Similar implementations of productive friction have been explored to promote creativity and reflection in programming education ~\cite{jonsson2022cracking, Kazemitabaar2025Exploring}.

\subsection{\textcolor{teal}{Towards} Instilling Designerly Thinking through Productive Friction} 

Prior research suggests that exposure to high-fidelity AI images causes design fixation~\cite{Wadinambiarachchi2024TheThinking}. In contrast, \citet{Cheng2014ADesigners} have demonstrated that ambiguous imagery can disrupt the ``path of least resistance'' and reduce fixation. Recent studies such as Inkspire~\cite{Lin2025Inkspire:Sketching} and Reframer~\cite{Lawton2023DrawingCo-CreativenbspAI}, both of which use sketching as an interaction medium, characterise unexpected AI outputs as serendipitous. This aligns with the literature on design by analogy~\cite{Xu2021DifferencesDesign-by-analogy}, which suggests that misinterpretations can act as creative stimuli leading to novel ideas~\cite{Casakin1999ExpertiseEducation}.

Therefore, to encourage divergence~\cite{Cheng2014ADesigners}, we introduced sketching and deliberately restricted \texttt{SketchifAI} to produce imperfect, low-to-mid-fidelity outputs. For instance, when participant P02 sketched a snail, \texttt{SketchifAI} reinterpreted the spiral shell as a stylised horn and the body as a hybrid of tortoise and bird, morphing the snail into a bipedal character (see Figure~\ref{fig:AIMis}), a transformation with the potential to inspire mascot characters, \textcolor{teal}{illustrating how productive friction may prompt unexpected creative directions}. 

While this resistance was intentional, it created a disconnect for users unaware of the constraint. Some participants perceived this as a lack of state-of-the-art capabilities, and interpreted unexpected outputs not as creative sparks, but as ``technical glitches.'' This mismatch between pedagogical intent (using low-to-mid fidelity to prevent fixation) and user expectations (of immediate, high-fidelity output) likely contributed to lower enjoyment scores. Without explicit scaffolding, productive friction \textcolor{teal}{can be} easily misconstrued as a system error. It is important to note that the rationale behind this friction (implemented via mid-fidelity outcomes) was withheld until the debrief in order to observe authentic user reactions. \textcolor{teal}{These preliminary findings suggest} a critical caveat: intentionally making AI outputs low-fidelity (rough or messy) to induce productive friction \textcolor{teal}{may be less effective} if the pedagogical reasoning is opaque to the user.

Therefore, we \textcolor{teal}{propose that embedding productive friction into the system architecture alone is insufficient; explicitly communicating the pedagogical rationale to the user may be required. Intentional friction may be more effective when transparent, and future interfaces might benefit from explicitly labelling resistance mechanisms} (e.g., \textit{``Ambiguity Mode Active''}) to nudge user frustration towards reflective practice.

\subsection{Designing Tools for Dialogic Designer-AI Interactions}

Recent work by \citet{Lee2024TheProcess} proposed that text supports divergent exploration, while sketching supports convergent refinement. Our findings suggest a more nuanced relationship; specifically, we observed that the \texttt{Sketch} condition \textcolor{teal}{showed a trend towards enhancing} \textsc{Fluency}, a key aspect of divergent thinking. We interpret this to mean that when the sketch interface did not provide an immediate, polished solution, users were denied a cognitive shortcut. Once participants began to sketch, it became cognitively more efficient to iterate visually than to stop, switch contexts, and formulate a text prompt. Consistent with Sch{\"o}n's reflection-in-action \cite{schon2017reflective}, the ambiguity of the medium may have compelled users to think through the sketching, reinforcing Goldschmidt's view of sketching as a fundamental thinking tool~\cite{Goldschmidt2017ManualRelevant}.

In addition, the \texttt{SketchPlusTags} modality, which mitigated negative effects on expressing intent, actually reduced the Originality of the products. We attribute this to a \textit{Semantic Anchor effect}. In the Text-only condition, the AI had the freedom to hallucinate visual details within semantic bounds. In the Sketch-only condition, it interpreted semantics within visual bounds. However, when providing both a specific visual structure (sketch) and a rigid semantic definition (tags), participants inadvertently over-constrained the generative model (see Figure \ref{fig:SketchPlusTags}). It is also possible that cognitive offloading played a role: the presence of textual tags may have encouraged users to produce simpler, less detailed sketches, relying on the text to \textit{fill-in-the-gaps}, which resulted in more generic outputs.

This finding complicates Shneiderman’s ideal of ``\textit{Low Threshold, High Ceiling}'' interfaces in creativity support tools \cite{Shneiderman2007CreativityInnovation} and prompts a rethink of how we design for AI-powered creativity. While the \texttt{SketchPlusTags} hybrid mode successfully lowered the threshold (restoring usability), the semantic anchors simultaneously lowered the ceiling for creativity. \textcolor{teal}{This highlights a potential trade-off: the interface that felt the safest (\texttt{SketchPlusTags}) resulted in a low threshold, low ceiling environment, potentially limiting unique ideas.}

Some participants treated the AI as a ``black box''~\cite{murray2022metaphors}, leading to trial-and-error frustration. They expressed a desire for the AI to seek clarification, asking, ``Is this what you mean?'' Without the ability to explain the intent behind a specific shape or correct the AI’s misunderstandings in real time, \texttt{Sketch} Interface felt like a one-way command rather than a co-creative partnership. To address this, we propose shifting to mixed-initiative co-creative interfaces~\cite{Deterding2017Mixed-InitiativeInterfaces}.

We suggest that future tools incorporate a conversational layer that goes beyond simple chat. For example, the AI could pause to query intent (e.g., ``I see a circular shape. Is this a face or a wheel?'') before rendering. Furthermore, the AI could provide reasoning for its creative choices. \textcolor{teal}{This conversational grounding could break the silent validation loop}, transforming it into a constructive dialogue that builds mutual understanding. This dialogue need not be limited to text but can leverage visual interactions, such as:

\begin{itemize}[leftmargin=*, topsep=5pt]
    \item Highlighting specific regions of ambiguity.
    \item Extending sketches to propose completions or include analogous ideas.
    \item Ghost interactions (faint overlays) that visualise potential design directions before they are finalised.
\end{itemize}

In addition, interfaces could also support non-linear exploration. By maintaining a version history similar to Git\footnote{https://git-scm.com}, tools can enable participants to branch ideas and toggle between problem-framing and solution-finding without the fear of losing previous ideas.

Finally, we propose that future AI design tools be task-adaptable, prompting users to reason about their modality choices. If an interface detects that a user repeatedly iterates on the same idea, the AI could suggest: ``You seem to be refining details. Would you like to switch to text?'' Such meta-cognitive prompting can help students learn to deploy modalities as distinct cognitive tools.

\subsection{Limitations}

\textcolor{teal}{We acknowledge several limitations that should be considered when interpreting our findings.} 

\textit{First}, our sample size was small ($N=9$) \textcolor{teal}{which is a major limitation for deriving stable quantitative findings}. For this reason, we employed Bayesian analysis, which is particularly well-suited for small-sample studies, as it provides probabilistic estimates of effects without relying on the large-sample assumptions required by frequentist statistics. \textcolor{teal}{However, the posteriors carry substantial uncertainty and estimates may shift with additional data. We therefore acknowledge that the quantitative findings should be treated as preliminary, and future studies with larger samples are required.} In addition, our participants were drawn from a single institution - although, one with a diverse student cohort. Our findings should be interpreted as signals of emerging behaviours in ``AI-native'' design students rather than as indicative of the entire population. 

\textit{Second}, our study utilised a rapid ideation task, and therefore can support only initial insights; we recognise longitudinal research is needed to evaluate sustained tool appropriation. 

\textit{Third,} we employed ControlNet, which requires semantic prompts to interpret sketch outlines -- we used a back-end prompt to maintain the scope and implement productive friction. We acknowledge that AI-models with native image-recognition capabilities may function differently; thus future research should evaluate how evolving architectures impact the intentional implementation of friction in creative workflows.

\section{Conclusion}

Through this study, we investigated how design students communicate intent to AI via \texttt{Text}, \texttt{Sketches}, and \texttt{SketchPlusTags} modalities, and how they affect creativity support and divergent thinking. Our findings \textcolor{teal}{point to} a critical paradox: While it is intuitive to assume that design students are visual thinkers, in practice they gravitated toward text-prompts. However, our \textcolor{teal}{preliminary evidence suggests that} the productive friction of sketching \textcolor{teal}{may increase Fluency, whereas the hybrid \texttt{SketchPlusTags} modality appeared to create a \textit{semantic anchor} limiting the Originality of generated products}. \textcolor{teal}{Taken together, these preliminary findings suggest that the challenge for the HCI and Design communities is not simply to maximise efficiency, but to develop AI-CSTs that promote reflection-in-action. We must explore} transitioning from making black-box content generators towards co-creative partners that provoke reflective thought, ensuring that AI-driven speed does not come at the cost of the design students' cognition.

\begin{acks}
This research is supported by Melbourne Research Scholarship, the Diane Lemaire Scholarship and the Rowden White Scholarship offered by the University of Melbourne. We would also like to thank Oshan Wisumperuma for his assistance with the back-end development of \texttt{SketchifAI}, and Jeremy Silver at the Melbourne Statistical Consulting Platform for their support.

\end{acks}

\bibliographystyle{ACM-Reference-Format}
\bibliography{paperbib}

\newpage

\appendix

\section*{Appendix}

\setcounter{table}{0}

\section{\textcolor{teal}{Researchers' Positionality}}
\label{sec:position}
\textcolor{teal}{
The first author has over two years of professional experience in UX/UI design and holds a bachelor's degree in design, alongside over three years of experience teaching UX/UI design and conducting HCI research at university level. She conducted the interviews and performed the initial coding and theme development. This background enabled her to empathise with participants' accounts of design practice, while remaining attentive to the risk of over-identifying with participant experiences given her shared background in design. The second and third authors contributed to the iterative team discussions, theme refinement, and finalising the analysis. Both hold doctorates and have over 20 years of combined experience teaching and researching UX/UI design. The collective expertise of the authorship team facilitated critical reflection on participants' accounts and contributed to constructing meaning from the data beyond surface-level description.
}

\section{\textcolor{teal}{Bayes Factor Thresholds}}
\renewcommand{\thetable}{B}
\label{tab:bf}
\begin{table}[ht]
    \centering
    \caption{Classification scheme for the interpretation of $\ln(BF)$,adapted from~\citet{wagenmakers2011psychologists}. The $\ln(BF)$ column shows Bayes factor thresholds converted to the natural log scale. Positive values indicate evidence in favour of the hypothesised direction (e.g., Sketch > Text), negative values indicate evidence in favour of the opposite direction (e.g., Sketch < Text) , and values near zero indicate absence of evidence for either hypothesis. The raw $(BF)$ column is provided for reference. Classifications are heuristic and should not be treated as hard thresholds. Here $H_1$ refers to the hypothesised direction (e.g., Sketch > Text) and $H_0$ refers to the opposing direction (e.g, Sketch < Text).}
    \small{
    \begin{tabular}{>{\centering\arraybackslash}m{2cm} >{\centering\arraybackslash}m{1.5cm} m{4cm} }
        \toprule
        \textbf{ln($BF$)} & \textbf{$BF$} & \textbf{Evidence category} \\ \midrule
        $> 4.61$      & $> 100$         & Extreme evidence for $H_1$ \\ 
        $3.40 - 4.61$ & $30 - 100$      & Very strong evidence for $H_1$ \\ 
        $2.30 - 3.40$ & $10 - 30$       & Strong evidence for $H_1$ \\ 
        $1.10 - 2.30$ & $3 - 10$        & Moderate evidence for $H_1$ \\ 
        $0 - 1.10$    & $1 - 3$         & Anecdotal evidence for $H_1$ \\ 
        $0$           & $1$             & No evidence \\ 
        $-1.10 - 0$   & $1/3 - 1$       & Anecdotal evidence for $H_0$ \\ 
        $-2.30 - -1.10$ & $1/10 - 1/3$  & Moderate evidence for $H_0$ \\ 
        $-3.40 - -2.30$ & $1/30 - 1/10$ & Strong evidence for $H_0$ \\
        $-4.61 - -3.40$ & $1/100 - 1/30$ & Very strong evidence for $H_0$ \\ 
        $< -4.61$     & $< 1/100$       & Extreme evidence for $H_0$ \\ \bottomrule
    \end{tabular}
    }
\end{table}

\newpage
\section{Design Briefs}
\label{appendix_B}

\subsubsection*{Design Brief - A }

Read the instructions given below and come up with as many design ideas as possible for a mascot character.

\noindent \textbf{Project overview:}	Design a mascot character to serve as a key visual element and the logo for an app designed for older adults to maintain their social connectedness with family and loved ones who live remotely.

\noindent \textbf{Purpose of the character:}	The image will represent the character offering comfort, guidance, and emotional connection. It should make the mascot feel more personal and approachable.

\noindent \textbf{Target audience:}	Individuals aged 65 or older.
\noindent \textbf{Things to consider:}	The mascot will appear in the logo, loading screens, onboarding flows, and occasional UI interactions.

\noindent Be creative and come up with as many ideas as possible, and sketch on the provided papers. Remember, you can annotate the sketch if you need to explain more about your design. Also, please remember always to give a number for each sketch you draw.

\subsubsection*{Design Brief - B}

Read the instructions given below and come up with as many design ideas as possible for a mascot character.
\noindent \textbf{Project overview:}	Design a mascot character to serve as a key visual element and the logo for a nature exploration and hiking app that promotes eco-friendly outdoor exploration and environmental awareness.

\noindent \textbf{Purpose of the character:}	The mascot should embody an element of nature, a sense of adventure, and a gentle guiding presence. It will support users on trails, share conservation tips, and celebrate sustainable actions like picking up litter or choosing low-impact routes.

\noindent \textbf{Target audience:}	Individuals aged 18-44, who are more likely to prioritise sustainability when making travel decisions.

\noindent \textbf{Things to consider:}	The mascot will appear in the logo, loading screens, onboarding flows, and occasional UI interactions.

\noindent Be creative and come up with as many ideas as possible, and sketch on the provided papers. Remember, you can annotate the sketch if you need to explain more about your design. Also, please remember always to give a number for each sketch you draw.
 
\subsubsection*{Design Brief - C}

Read the instructions given below and come up with as many design ideas as possible for a mascot character.
\noindent \textbf{Project overview:}	Design a mascot character to serve as a key visual element and the logo for a music creation app that contains inspirational tones from folk music traditions. 

\noindent \textbf{Purpose of the character:}	The character should reflect the app’s mission to honour cultural roots while promoting creativity and modern expression. It must feel soulful, respectful, and vibrant, bridging tradition and technology.

\noindent \textbf{Target audience:}	Individuals those aged 16 or above, who love to create music while honouring and respecting culture and traditions.

\noindent \textbf{Things to consider:}	The mascot will appear in the logo, loading screens, onboarding flows, and occasional UI interactions.

\noindent Be creative and come up with as many ideas as possible, and sketch on the provided papers. Remember, you can annotate the sketch if you need to explain more about your design. Also, please remember always to give a number for each sketch you draw.

\onecolumn

\clearpage
\section{Participant Information}
\renewcommand{\thetable}{D}
\label{appendix A}
\begin{table}[ht] 
\centering
\caption{Participant information including participant id, degree level, major, year 
of study, AI tools used, self-reported AI experience (AI Exp.), 
prompting confidence (Prompt. Conf.), and sketching confidence 
(Sketch. Conf.), each rated out of 10, age, and gender.}
\begin{tabular}{p{0.25cm} p{0.75cm} p{1.5cm} l p{3.6cm} p{0.5cm} p{1.05cm} p{0.75cm}p{0.5cm} l}
\hline
\textbf{PID} & \textbf{Degree} & \textbf{Major} &\textbf{Yr.} & \textbf{AI tools used} & \textbf{AI Exp.} & \textbf{Prompt. Conf. (/10)} & \textbf{Sketch. Conf. (/10)} & \textbf{Age} & \textbf{Gender} \\
\hline \\
P01 & B.Des  & UX Des.~\& Comp. & 3 & ChatGPT, Figma AI, Perplexity, Google Gemini & 1 & 7 & 6 & 22 & M \\ \\
P02 & B.Des  & UX Des.~\& Perform. Des. & 1 & ChatGPT, CanvaAI, AdobeAI & 1 & 7 & 8 & 19 & F \\ \\
P03 & B.Des  & UX Des. & 1 & Cursor, MagicAnimator, ChatGPT & 1 & 6 & 8 & 20 & F  \\ \\
P04 & M.Des  & Arch.~\& Graphics & 2 & ChatGPT, Midjourney, Finch & 1 & 6 & 8 & 28 & M  \\ \\
P05 & B.Des  & UX Des.~\& Comp. & 1 & Midjourney, Canva, ChatGPT, Adobe AI, Grok & 2 & 7 & 8 & 19 & F \\ \\
P06 & B.Des  & UX Des. & 2 & ChatGPT, Framer, Figma, Adobe, Canva & 2 & 7 & 5 & 19 & F \\ \\
P07 & B.Des  & UX Des. & 1 & ChatGPT, Dall-E & 2 & 8 & 5 & 20 & M \\ \\
P08 & B.Des  & Humanit.~\& UX Des. & 3 & Mid journey, Sora, ChatGPT, Adobe Firefly & 0.5 & 5 & 4 & 21 & M \\ \\
P09 & B.Des  & UX Des. & 2 & ChatGPT, Canva AI & 1 & 6 & 4 & 21 &  F \\ \\
\bottomrule
\end{tabular}
\end{table}

\end{document}